\documentclass[prb,aps,twocolumn,groupedaddress,floats,showpacs,final, fleqn]{revtex4-1}
\usepackage{dcolumn}
\usepackage{bm}
\usepackage{color}
\usepackage{mathtools}
\usepackage{amsmath,amsfonts,amssymb,bm}
\usepackage{graphicx,tikz}
\usepackage[pdftex, bookmarks=true, linktoc=section, pdfstartview=FitH]{hyperref} 

\begin{document}

\author{Olga Goulko}
\affiliation{Department of Physics, University of Massachusetts, Amherst, MA 01003, USA}
\affiliation{Present address: Raymond and Beverly Sackler School of Chemistry and School of Physics and Astronomy, Tel Aviv University, Tel Aviv 6997801, Israel}
\author{Nikolay Prokof'ev}
\affiliation{Department of Physics, University of Massachusetts, Amherst, MA 01003, USA}
\affiliation{National Research Center ``Kurchatov Institute,"
123182 Moscow, Russia}
\author{Boris Svistunov}
\affiliation{Department of Physics, University of Massachusetts,
Amherst, MA 01003, USA}
\affiliation{National Research Center ``Kurchatov Institute,"
123182 Moscow, Russia}
\affiliation{Wilczek Quantum Center, School of Physics and Astronomy and T. D. Lee Institute, Shanghai Jiao Tong University, Shanghai 200240, China}

\title{Restoring a smooth function from its noisy integrals}
\date{\today}

\begin{abstract}
Numerical (and experimental) data analysis often requires the restoration of a smooth function from a set of sampled integrals over finite bins. We present the bin hierarchy method that efficiently computes the maximally smooth function from the sampled integrals using essentially all the information contained in the data. We perform extensive tests with different classes of functions and levels of data quality, including Monte Carlo data suffering from a severe sign problem and physical data for the Green's function of the Fr\"ohlich polaron. \end{abstract}


\maketitle

\relpenalty=9999
\binoppenalty=9999

\section{Introduction}
\label{sec:introduction}

Sampling is at the heart of many Monte Carlo computations and experimental measurements. Data points are generated or measured and we eventually want to restore the underlying smooth probability density distribution $f(x)$ behind them. Many density estimation protocols for lists of statistical data exist,\cite{KDEbookNarskyPorter,KDEbookScott, KDEbookSilverman} but for large-scale Monte Carlo calculations storing the individual data points would require pentabytes of memory 
and lead to a substantial slowing down of the simulation. More importantly, the individual data points are not needed, provided that we know that $f(x)$ is structureless below a certain scale. Hence, without losing any essential information, we can collect integrals of $f(x)$ over finite-size bins. The problem then is how to extract all the information from these integrals without systematic bias, augmenting it with our {\it a priori} knowledge of the smoothness of the function.

In this paper, we propose a simple and efficient solution, which we call the bin hierarchy method (BHM).\cite{goulko2017bhmimplement} We note that the problem is closely related to the problem of numerical analytic continuation under consistent constraints,\cite{ACpaper} but with a crucial simplification. Unlike for the numerical restoration of spectral functions, we do not have to account for possible sharp features, such as $\delta$-function peaks or kinks, which cannot be resolved within the sampled error bars. Assuming that $f(x)$ is smooth, we are allowed to parametrize its optimal approximation $\tilde{f}(x)$ as a polynomial spline (or a spline with a general functional ansatz). 

The protocol of constructing $\tilde{f}(x)$ has some similarities with the smoothing spline approach.\cite{SmoothingSplinesDeBoor} However, there is a fundamental difference in the data structure. Rather than fitting to approximate function values on a given set of points, we fit the spline directly to the sampled integrals. The central requirement is that $\tilde{f}(x)$ must be consistent not only with the sampled integrals over the elementary histogram bins, but also with the integrals over any combination of these. Integrals over large bins are known with higher precision than those over smaller bins, since the former contain more data points. Hence, accurately recovering a large integral is a priority over recovering a smaller one. In particular, bins below a certain scale contain noise rather than information about $f(x)$, and thus can be safely omitted in the fitting. We formalize this idea by introducing a hierarchy of bins with one bin over the entire domain of $f(x)$ on the top level, and doubling the number of bins on each subsequent level. The goodness of fit is evaluated on each level $n$ separately, with $\chi_n^2/\tilde{n}$ as the measure, where $\tilde{n}$ is the number of bins on the given hierarchy level.

Out of all possible approximations $\tilde{f}(x)$ consistent with the sampled integrals we select the one with the least features, i.e.\ the one with the least fitting parameters. This is achieved by fitting a spline of order $m$ (typically $m=3$) with the minimal number of knots that yields an acceptable fit. The number and the positions of the knots are determined automatically by the fitting algorithm. Known boundary conditions on $f(x)$ can also be accounted for. If desired, additional optimization terms, for instance for the jump in the highest spline derivative, can be included into the fitting procedure, with iterative improvement in the spirit of the method of consistent constraints.\cite{AnalyticContConsConstr}

The paper is organized as follows. In Sec.~\ref{sec:regular}, we review alternative methods to construct $\tilde{f}(x)$ and argue that the BHM is superior in accuracy, efficiency, and simplicity. The details of the BHM are presented in Sec.~\ref{sec:ourmethod}. In Sec.~\ref{sec:tests}, we present several tests of the BHM. We conclude in Sec.~\ref{sec:conclusions}.

\section{Review of alternative methods}
\label{sec:regular}
\subsection{Naive histograms}
The most straightforward way to approximate a distribution $f(x)$ is the histogram:\cite{KDEbookNarskyPorter,KDEbookScott, KDEbookSilverman} divide the domain into several bins and count how many times the sampled values of $x$, which were generated with probabilities given by $f(x)$, fall into each of the bins. For each generated $x\in\textnormal{bin}_i$, the counter $c_i$ for the bin $i$ is increased by one. The distribution is then approximated by
\begin{equation}
\tilde{f}(x) = \sum_i \, \frac{c_i }{N \Delta_i}\, \Pi_i(x) ,
\label{eq:hist}
\end{equation}
where $N$ is the total number of sampled points, $\Delta_i$ is the width of the bin $i$, and $\Pi_i(x)$ is the boxcar function, which is one if $x\in {\rm bin}_i$ and zero otherwise. In what follows, we  refer to this method as ``naive histogramming"---to emphasize the contrast with the BHM that also involves histogramming as a part of the protocol. The end result of the naive histogram method is a staircase function, which is not smooth. This drawback can be ameliorated by taking $c_i/(N\Delta_i)$ as the value of $\tilde{f}(x)$ at the bin centers and fitting a smoothing spline to the data with their statistical error bars.\cite{SmoothingSplinesDeBoor} A more important issue is that the naive histogram method suffers from an inherent compromise between the resolution of features and statistical noise. Sufficiently narrow bins are necessary to resolve $f(x)$ without introducing a significant systematic bias, while reducing the size of the bins increases the noise in the sampled counters.

\subsection{Basis projection method}
A generalization of the naive histogram method---the basis projection method---can significantly reduce this drawback. This method is based on a generalized Fourier series expansion of $f(x)$. For each bin $i$, we choose a basis $\{ e_i^{(j)}(x)\}$ of $m_i$ real-valued functions with $e_i^{(j)}(x)=0$ if $x \notin {\rm bin}_i$ that satisfy the orthonormality condition
\begin{equation}
\langle e_i^{(j)}|e_i^{(j')}\rangle\equiv\int_{\textnormal{bin}_i}w_i(x)e_i^{(j)}(x)e_i^{(j')}(x)dx=\delta_{jj'}.
\label{eq:onb}
\end{equation}
The non-negative weight function $w_i(x)$ of the inner product should be chosen such that all integrals are convergent. If no divergences are present $w_i(x)$ can be set to unity. The basis and weight functions may be different for each bin, and can be defined also for bins of infinite size. In general, the basis functions can be chosen to reflect the properties of $f(x)$ that are known in advance, for instance known divergences or asymptotic behavior. Otherwise, in the majority of the cases, shifted Legendre polynomials are a reasonable choice. 

With the basis projection method, in each bin $i$ we keep track of several counters $c_i^{(j)}$ estimating the $m_i$ generalized Fourier coefficients $\langle e_i^{(j)} | f  \rangle$. Specifically, each time we generate $x\in\textnormal{bin}_i$ we add the value $w_i(x)e_i^{(j)}(x)$ (rather than one) to each of the $c_i^{(j)}$ counters. The function $\tilde{f}(x)$ is then given by
\begin{equation}
\tilde{f}(x) = \sum_{i} \sum_{j=1}^{m_i} \frac{c_i^{(j)}}{N} e_i^{(j)}(x).
\label{eq:basis}
\end{equation}
Note that Eq.~(\ref{eq:hist}) is a special case of the above formula with $m_i=1$, $w_i(x)=1$, and $e_i^{(1)}(x)=\Pi_i(x)/\sqrt{\Delta_i}$.

It can be easily seen that this procedure corresponds to the best possible representation of $f(x)$ in terms of orthogonal basis functions on each bin $i$, since minimizing
\begin{eqnarray}
\int_{\textnormal{bin}_i} w_i(x)(f(x)-\tilde{f}(x))^2dx=\nonumber\\
=\int_{\textnormal{bin}_i}w_i(x)\left(f(x)-\sum_{j=1}^{m_i} \frac{c_i^{(j)}}{N} e_i^{(j)}(x)\right)^2 dx
\end{eqnarray}
with respect to the $c_i^{(j)}$ implies
\begin{equation}
\int_{\textnormal{bin}_i} w_i(x)e_i^{(j)}(x)\left(f(x)-\sum_{j'=1}^{m_i} \frac{c_i^{(j')}}{N} e_i^{(j')}(x)\right) dx=0,
\end{equation}
and consequently $c_i^{(j)}/N=\langle e_i^{(j)} | f  \rangle$.

Using basis projections we collect more information in each sampling step than with the naive histogram, since some information about the position of $x$ inside the bin is preserved. Therefore fewer bins are necessary to resolve the function. The bin widths and basis sizes $m_i$ can be chosen in a way that the systematic error is negligible compared to the statistical error, without a substantial increase in the statistical noise of the $\tilde{f}(x)$ values.

While the basis projection method is a significant improvement over the naive histogram, it still has several disadvantages. The binning and the basis functions must be chosen at the beginning of the sampling process and cannot be altered retrospectively. Basis projections can also be computationally expensive, especially for a large basis, or a basis built of complicated functions. Moreover, the sampled projections generally exhibit large jumps at the bin boundaries because the basis functions may be unbounded. As before, we can eliminate these jumps by taking the values of $\tilde{f}(x)$ at several points inside each bin and then constructing a smoothing spline. This procedure, however, involves an unnecessary loss of information. Rather than reducing the fit to specific individual points, it is suggestive to fit a spline to the sampled integrals $\langle e_i^{(j)} | f  \rangle$ directly, resulting in smaller errors. In addition, we can extract more exhaustive information about $f(x)$ by using overlapping bins of different size that cover a range of relevant scales of the function. This is achieved by the BHM described in Sec.~\ref{sec:ourmethod}.

\subsection{Reweighting}
Reweighting is a Monte Carlo-specific technique that allows one to convert the statistics generated for a given variable $x$ to statistics for other values of this variable.\cite{samplingRW, sompolaron} This is achieved in the following way. Prior to sampling, we choose a fixed set of points $x_i$ and associate a bin with each of the $x_i$. The bins can be large and overlapping. Whenever the random variable $x$ falls into the bin $i$ associated with the point $x_i$, we update the $x_i$ counter by the reweighted value $vp(x_i)/p(x)$. Here $v$ is the value that would have been sampled without reweighting, and $p(x_i)/p(x)$ is the ratio of the probability weights at $x_i$ and $x$, where $p(x)$ is the probability distribution used to sample the physical function $f(x)$. This can be seen as a deterministic ``update" from the generated point $x$ to the fixed point $x_i$. For sufficiently large bins, several counters may be updated simultaneously for each generated $x$. The reweighting technique provides an unbiased estimator for the value $f(x_i)$ at any fixed $x_i$.

The reweighting method is very efficient in the case when one is interested in a single special value of a certain continuous variable. A characteristic example is the sampling of the single-particle density matrix from the statistics for the Green's function by reweighting the latter to the imaginary-time value $\tau = -0$. In all other cases, reweighting is computationally expensive compared to the gain in accuracy. It also requires a case-specific implementation that assumes knowledge of the analytical form of the reweighting factors. To be maximally efficient, one has to adjust the size of ${\rm bin}_i$ for each $x_i$ of interest. The bin size must be optimized in such a way that, on the one hand, it is large enough to ensure sufficiently large statistics, and on the other hand small enough that the reweighting factors remain of the order of unity (to avoid numerous vanishingly small contributions that consume simulation time). Bins that are too large can also result in a strong increase in the autocorrelation time, since reweighting from a rare event to a frequent one implies a large $p(x_i)/p(x)$ ratio, which will result in occasional anomalously large contributions to the sampled counters. Moreover, to produce a smooth outcome, the reweighting protocol requires an appropriately dense mesh of points with strongly overlapping bins, which is computationally expensive since a large number of counters needs to be updated for each generated $x$. For a continuous set of function values, a smoothing spline needs to be fitted at additional computational cost, comparable to the cost of the BHM fit described in Sec.~\ref{sec:ourmethod}.

\section{Bin hierarchy method}
\label{sec:ourmethod}

The BHM restores smooth distributions in a universal, unbiased, and efficient manner. The method satisfies the following requirements:
\begin{itemize}
\item All the information contained in the sampled data should be used to fit the distribution: the accuracy should be essentially the same as if we had stored the full list of the generated values of $x$. This is achieved by introducing a hierarchy of overlapping histogram bins of different size, where large bins give precise estimates of the broad features of the distribution, while small bins resolve its fine structure once enough data has accumulated. Note that we assume that the sampled distribution is structureless below a certain scale, and hence there always exists a binning that is sufficiently fine to resolve all of its features.
\item The resulting function should be exactly smooth, meaning that the function and all its derivatives up to a given order must have no jumps. This is achieved by fitting a polynomial (or generalized) spline.
\item Out of all smooth functions consistent with the sampled data within its error bars, the selected result should have the least features (peaks, oscillations, etc.). This is  achieved by choosing the spline with the minimal number of free parameters that fits the data.
\item The sampling process and the restoration of the smooth function should be computationally and memory efficient. In particular, they should be more efficient than the basis projection method.
\item The algorithm should work well regardless of the quality and the number of sampled data. As more data are collected and statistics improves, so should the fitted distribution without needing any external adjustments.
\item The algorithm should be highly automated. No human input or control should be necessary other than a limited choice of initial parameters. In particular, the method should be applicable to all reasonably occurring functions and should require no adjustment for different classes of functions.
\end{itemize}
Below we describe the setup in detail.

\subsection{Sampling stage}

We divide the domain into $2^K$ non-overlapping elementary bins that do not need to be of equal width. Because in the end we form combinations of several bins, the elementary bins may be arbitrarily small. Bins with too few data points are automatically excluded from the analysis. As the number of sampled points increases, bins of smaller size will become usable. The number of elementary bins is thus limited only by memory. The $K\rightarrow\infty$ limit would formally correspond to keeping the full list of the generated values of $x$. In practice, the distributions of interest are structureless beyond a certain scale that sets a natural limit on the required elementary bin width.

We generate values of $x$ according to a probability distribution given by $|f(x)|$ (importance sampling) and sample the value $v=\textnormal{sign}[f(x)]$ in the elementary bin $i$ that contains $x$. We can also include additional weighting coefficients into the sampled values. For each bin we store the number of sampled values in the bin, $N_i$, as well as the average $\bar{v}_i$ and the scaled variance $M_2(v_i)=(N_i-1)$Var$(v_i)$ of the sampled values over the bin. We obtain the sampled integral $I_i$ over each elementary bin via $I_i=\bar{v}_i N_i/N$, where $N$ is the total number of sampled points. The variance of the sampled integral is calculated via
\begin{eqnarray}
{\rm Var}(I_i)&=&\frac{M_2(I_i)}{N-1}\\
M_2(I_i)&=&M_2(v_i) \, +\, \bar{v}_i^2 \, \frac{N_i(N-N_i)}{N}
\end{eqnarray}
and its error is given by $\delta I_i =\sqrt{{\rm Var}(I_i)/N}$.

\subsection{Bin hierarchy}

After sampling has finished, we construct combinations of elementary bins. These make up a hierarchy with $2^n$ bins at each level $n\in\{0,\ldots,K\}$. At the top level we have one large bin containing the entire integral; on the next level we have two bins containing the integrals over half of the interval each, etc. The fit is constructed to minimize
\begin{equation}
\sum_{n=0}^K\frac{\chi^2_n}{2^n} =\frac{\chi^2_0}{1}+\frac{\chi^2_1}{2}+\ldots + \frac{\chi^2_K}{2^K}\, ,
\label{eq:chisqmin}
\end{equation}
where $\chi^2_n$ is the goodness of the fit considering only the bins on the level $n$. Note that the construction of the bin hierarchy happens in the post processing. During the sampling stage, only the values $N_i$, $\bar{v}_i$ and $M_2(v_i)$ in the elementary bins (corresponding to the hierarchy level $K$) need to be collected and stored.

We tested several modifications of this setup, which produced consistent results. Including more than $2^n$ bins on each level by using overlapping bins of equal size was  found to bring no significant improvement. Likewise, using weights other than $1/2^n$ for $\chi^2_n$ did not improve the result. In general, finer bins at higher levels do not substantially contribute to the form of the fit, as their error bars are large and the shape of the distribution is usually already determined by the lower-level integrals. But they are still included into the fitting procedure (provided that they contain enough data points for statistics) to resolve potential fine structures of the distribution. For example, a periodically oscillating function like the sine can have zero average over large parts of the domain, while being nonzero on a finer scale. The structure of such a function will be resolved from the integrals over small bins if they contain enough data. Due to the $1/2^n$ weight factors, the small bins cannot overpower the large bin contributions.

Note that while for minimization we use the sum over all levels, to decide whether a fit is accepted we check the goodness of the fit at every level separately. A fit is accepted only if {\it each} of the $\chi^2_n/\tilde{n}$ is approximately 1, within a given number $T$ of standard deviations $\sigma=\sqrt{2/\tilde{n}}$ ($\tilde{n}\leq2^n$ is the number of bins on level $n$ which contain sufficient data to be used for fitting). The fit acceptance threshold $T$ is an external input parameter, typically $T=2$. Lower values of the threshold are likely to result in the failure to produce an acceptable fit. Instead of a fixed value, a range of $T$ can also be specified, for example between two and four. If there is no acceptable fit at the lowest threshold value, it is gradually increased until either the maximum is reached or an acceptable fit is found. For a large number of hierarchy levels, the statistical probability for a level to exceed a $2\sigma$ threshold on $\chi^2_n/\tilde{n}$ becomes non-negligible, even if the overall fit is good. Specifying a threshold range allows one to attempt fits with lower threshold values, without risking a failed fit.

\subsection{Polynomial fit}
The sampled integrals are fitted to a spline. In what follows we discuss polynomial (typically, cubic) splines, but a generalization to other functions is straightforward. Here we briefly review the theory of fitting a single polynomial $p(x)=\sum_{k=0}^m a_kx^k$, before we turn to the full spline in the next section.

There exists a lot of standard software for fitting a set of points to a function. Here we want to match the polynomial to a set of integrals, which requires a slight adjustment. 
The derivation follows the steps of general linear least squares fitting.\cite{NR} The best polynomial fit minimizes
\begin{equation}
\chi^2=\sum_i\left(\frac{I_i-I_i^{(p)}}{\delta I_i}\right)^2,
\label{eq:chisq}
\end{equation}
where $I_i^{(p)}$ is the integral of the polynomial over the bin $i$. For simplicity we omit the sum over bin levels and the corresponding weighting factors in this section, but the generalization is straightforward. Let us label the boundaries of bin $i$ by $\bar{x}_i$ and $\bar{x}_{i+1}$. The integral of the polynomial over the bin equals
\begin{equation}
\int_{\bar{x}_i}^{\bar{x}_{i+1}} p(x)dx = \sum_{k=0}^{m} \frac{a_k}{k+1}(\bar{x}_{i+1}^{k+1}-\bar{x}_i^{k+1}) \equiv \sum_{k=0}^{m} a_k X_{ik}
\end{equation}
and Eq.~(\ref{eq:chisq}) becomes
\begin{eqnarray}
\chi^2&=&\sum_i\left(\frac{I_i}{\delta I_i}-\sum_{k=0}^{m} a_k \frac{X_{ik}}{\delta I_i}\right)^2 \nonumber\\
&\equiv& \sum_i \left(b_i - \sum_k A_{ik}a_k\right)^2=|\mathbf{A}\cdot\mathbf{a}-\mathbf{b}|^2.
\end{eqnarray}
The vector $\mathbf{b}$ contains the sampled integrals divided by the error bar (its length equals the number of bins $\tilde{n}$), the vector $\mathbf{a}$ of length $m+1$ contains the free parameters of the polynomial, and the $\tilde{n}\times (m+1)$ design matrix $\mathbf{A}$ is defined by $A_{ik}=X_{ik}/\delta I_i$. The design matrix has more rows than columns since there must be more data points than fit parameters. We find the vector $\mathbf{a}$ that minimizes $|\mathbf{A}\cdot\mathbf{a}-\mathbf{b}|^2$ using singular value decomposition of the design matrix.\cite{NR} The error on the fitted polynomial at a given point $x$ is obtained from the covariance matrix $C_{jk}=\textnormal{Cov}(a_j,a_k)$ via
\begin{equation}
\delta p(x) = \sqrt{\sum_{i,j=0}^{m} C_{ij}x^{i+j}}.
\label{eq:coverror}
\end{equation}

\subsection{Spline}
At each knot between two spline pieces, the values of the corresponding polynomials and their derivatives up to order $(m-1)$ are equal. This gives $m$ constraints at each knot. Since each polynomial has $m+1$ coefficients, the total number of free parameters is $n_p+m$, where $n_p$ is the number of spline pieces. The extra $m$ parameters arise since there is one more polynomial piece than knots. Known boundary conditions, if any, further reduce the number of free parameters.

The algorithm determines the positions of the knots using the following procedure. Begin by attempting to fit one polynomial on the whole interval. If this fit is not acceptable, split the interval into two equal parts and attempt fitting a two-piece spline. If this fit is not acceptable either, check the goodness of the fit on each interval separately. Intervals where the fit is not acceptable have to be split in the next iteration, while all other intervals remain unchanged. Repeat this procedure until either an acceptable fit is found, or the intervals can no longer be split. We require that the set of equations for each spline piece be overdetermined, i.e.\ that the number of bins inside each interval is larger than $m+1$. This sets the limit on the maximal number of spline pieces.

To check the goodness of the fit on an individual interval, we compute the values of $\chi^2_n/\tilde{n}$, where $\chi^2_n$ is evaluated only on bins of level $n$ that are fully inside the given interval (and that contain enough statistical data), and $\tilde{n}$ is the number of such bins. Larger bins that reach over several intervals cannot be used for this check, but they are used to check the goodness of the overall fit. In all test cases, if the polynomials on the individual intervals passed the goodness-of-fit test, the overall spline evaluated over the entire domain passed also. Note that as before, the checks on the individual intervals are performed for each hierarchy level separately. If more than half of the level $n$ bins inside a given interval do not contain enough data, the check for this interval terminates without proceeding to subsequent levels.

\subsection{Constraining jumps in the highest derivative}
An additional constraint on the jumps in the $m$th derivative of the spline can be imposed on the final fit, if desired. Derivatives up to order $m-1$ are matched at the spline knots, while all derivatives of order greater than $m$ are zero. The piecewise constant $m$th derivative, which is proportional to the value of the parameter $a_m^{(j)}$ on each interval $j$, is the only one that exhibits jumps at the knots. Imposing an additional constraint that aims to minimize these jumps ensures a continuous evolution of the fitted spline as a function of the number of sampled points $N$. The number of spline pieces for the best fit generally increases with growing $N$, as better statistics permits a better resolution of the sampled function. Additional interval divisions are triggered at certain discrete values of $N$, namely when one of the $\chi^2_n$ values on the affected interval crosses the goodness-of-fit threshold. This creates an additional jump in the $m$th derivative at the position of the new knot. This jump can be closed with only a small sacrifice in $\chi^2_n$ since the previous spline with one fewer knot was almost acceptable. In this sense, the additional constraint on the jump acts as a compensation mechanism for a potentially too vigorous interval splitting. In most cases, constraining the jump has almost no effect of the fit.

The constraint term has the form
\begin{equation}
\frac{\lambda}{n_p}\sum_{j=1}^{n_p-1}\lambda_j\left(\frac{a_m^{(j+1)}-a_m^{(j)}}{\bar{a}_m^{(j+1)}-\bar{a}_m^{(j)}}\right)^2,
\label{eq:jumpconstraint}
\end{equation}
where $\bar{a}_m^{(j)}$ are the polynomial parameters of the optimal fit without the constraint and the weights $\lambda$ and $\lambda_j$ control the strength of the constraint. After computing the optimal spline through minimization of Eq.~\eqref{eq:chisqmin}, a new spline fit is attempted on the same interval division, this time by minimizing the sum of \eqref{eq:chisqmin} and the constraint \eqref{eq:jumpconstraint}. The local weights $\lambda_j$ are determined by the goodness of the original fit (without constraint) on the intervals $j-1$, $j$, $j+1$ and $j+2$, which are adjacent or next-to-adjacent to the respective knot. Specifically, we determine (on each hierarchy level of each of these intervals) the difference between the maximally acceptable $\chi^2_n/\tilde{n}=1+T\sqrt{2/\tilde{n}}$ and the actual $\chi^2_n/\tilde{n}$. We take $\lambda_j$ to be the minimum of all these differences. This ensures that the constraint does not affect spline intervals where already the original fit was barely acceptable. The global weight $\lambda$ is iteratively adjusted to the maximal value that is still compatible with the imposed threshold on the $\chi^2_n/\tilde{n}$. Additional iterations can be performed by replacing the $\bar{a}_m^{(j)}$ with the values calculated with the constraint.

\subsection{Errors on the spline coefficients}
An estimate on the spline coefficient errors can be obtained from Eq.~\eqref{eq:coverror} for each spline piece. This equation is accurate for parametric fits to a set of statistically independent normally distributed data points. This is not the case for the BHM, due to the overlapping bins and the unconventional $\chi^2$-minimization protocol. Because of this, one cannot attach Gaussian standard deviation probabilities to error bars. A robust way of defining the error at a given point $x$ is to look at the histogram of $\tilde{f}(x)$ values over a large set of independent runs with the same sample size $N$. The smallest interval around the histogram mean that contains $68.27\%$ of the $\tilde{f}(x)$ values corresponds to the robust estimate for two standard errors ($\pm1\sigma$). This is the most rigorous, albeit impractical, way of defining the error under these circumstances. A suitable alternative protocol should yield errors that are close to the robust value.

We performed extensive tests of several error estimation protocols by comparing them to the robust estimate for different types of distribution $f(x)$. These tests confirmed that Eq.~\eqref{eq:coverror} provides a good error estimate in most cases. In some cases the error was found to be overestimated by a factor of about two compared to the robust error. This occurred in particular at the domain boundaries and for BHM splines with a large number of knots. A concrete example is presented in Sec.~\ref{sec:osc}. The error was never observed to be too small and hence Eq.~\eqref{eq:coverror} gives a convenient and efficient way to obtain a conservative error estimate on the BHM fit.

A tighter error estimate can be obtained using bootstrap \cite{efron1979bootstrap} with the following protocol. Produce an array of $M$ histograms each of which contains a fraction $N/M$ of the $N$ sampled data points, so that each point is sampled in exactly one of the histograms. Generate $\tilde{M}\geq M$ bootstrap histograms by taking $\tilde{M}$ random linear combinations (with positive integer coefficients) of the $M$ histograms. Each of these bootstrapped histograms then contains $N$ sampled points, but with possible repetitions. Run the BHM fit on each of the bootstrap histograms fixing the positions of the knots to be the same as for the BHM fit of the regular histogram (the sum of all bootstrap histograms with unit weight) and without evaluating the goodness of fit. The error on the spline coefficients is then determined from the statistics on the bootstrapped histogram fits in the usual way. The bootstrap error was found to be very close to the robust error estimate in all our tests. This scheme uses more resources, especially memory, since $M\gtrsim100$ histograms need to be stored instead of one. The increase in computational time is less significant, since performing fits with fixed knot positions is very fast.

Another error estimation protocol is based on the analysis of the evolution of the BHM fit as a function of sample size $N$. For this, the BHM spline is produced and saved periodically in intervals of $\Delta$ sampling steps. The idea is that the fit results $\tilde{f}(x)$ will approach the exact value with the dispersion $\sigma\propto1/\sqrt{N}$. The interval size $\Delta$ must be large enough to guarantee that the contributions $A_k$ from different intervals are
statistically independent, allowing one to invoke the central limit theorem:
\begin{equation} 
\tilde{f}_k = \frac{1}{k}\sum_{i=1}^k A_i \Rightarrow A_k = k\tilde{f}_k - (k-1)\tilde{f}_{k-1}.
\end{equation}
The dispersion of $\tilde{f}_k$ can then be obtained from the dispersion $\sigma_*$ of the independent $A_k$ via $\sigma_k^2 = \sigma_*^2 /k$. An additional cutoff $k\geq k_0$ should also be introduced to reduce the effect of the initial portion of the statistics that can be strongly affected by transient processes. A simple consistency criterion is that there should be a range of $\Delta$ and $k_0$ such that $\sigma_*$ is essentially independent of both parameters. Our tests showed that the errors calculated with this method are almost indistinguishable from the ones obtained with bootstrap, and hence also agree with the robust error. This method is somewhat slower, since the full BHM fit needs to be evaluated repeatedly, and requires a sufficiently large $N$. On the other hand, the $A_k$ sequence provides explicit insight into the statistics of the data and can help to identify statistical fluctuations.

\begin{figure*}[htp]
\includegraphics[width=0.49\textwidth]{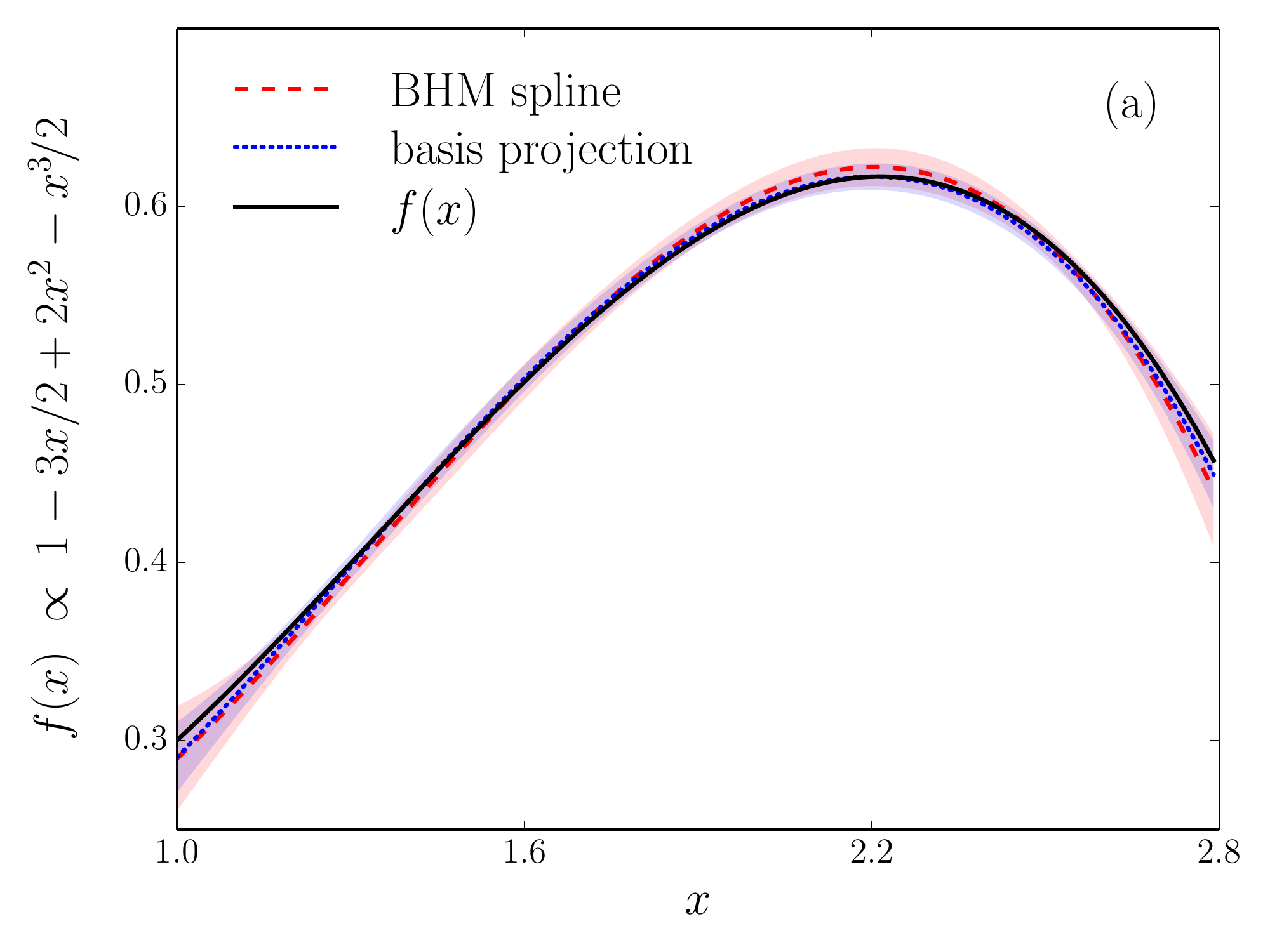}\hfill
\includegraphics[width=0.49\textwidth]{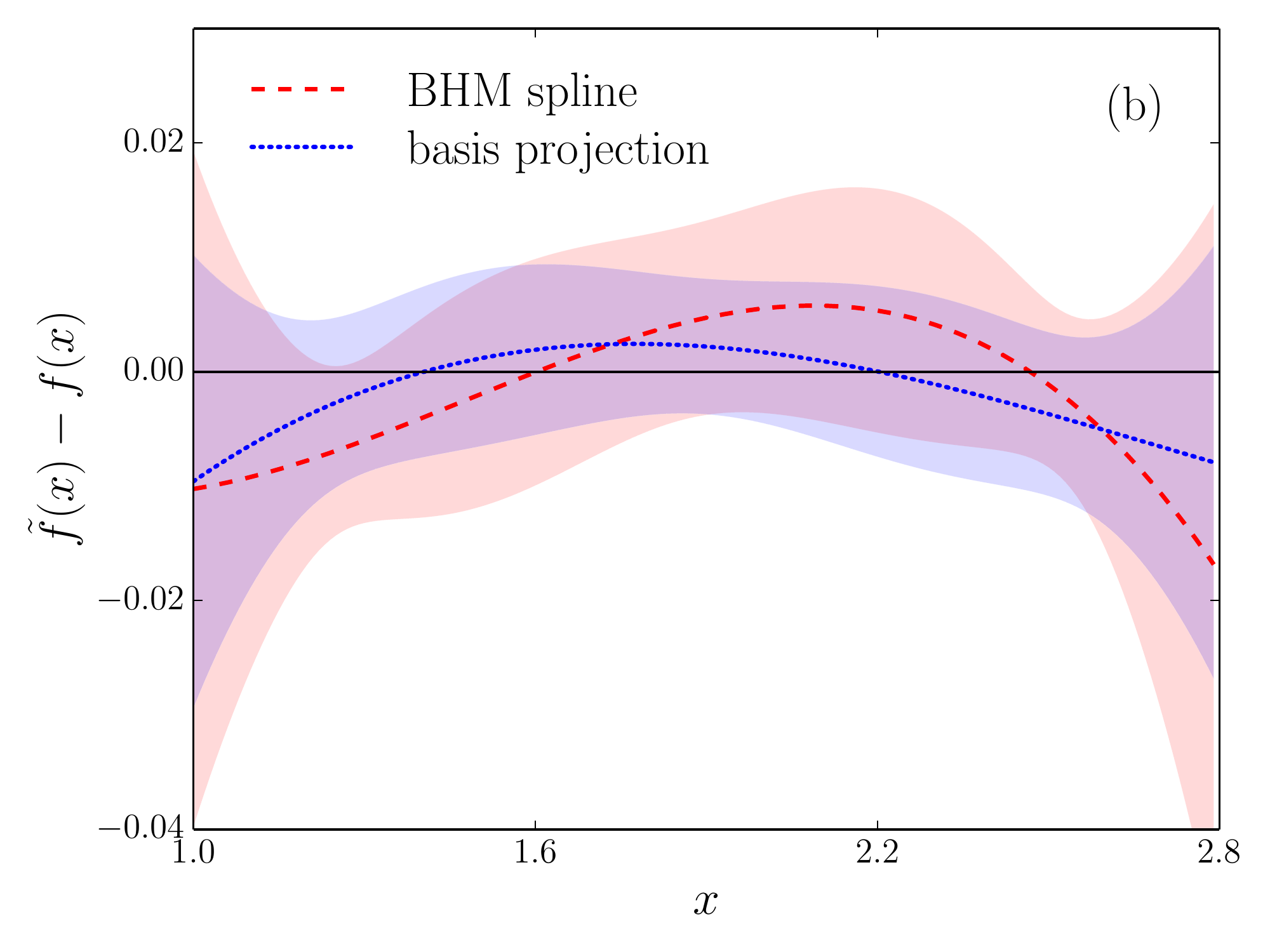}
\caption{\label{fig:poltest}Cubic polynomial test function. The BHM fit (red dashed line with error band) in comparison with the basis projection method (blue dotted line with error band). The left panel (a) shows the function and the fits, while the right panel (b) shows the difference between the fitted and the true function for both methods.}
\end{figure*}
\subsection{Sign problem}
\label{sec:signproblem}
Monte Carlo sampling often suffers from a sign problem, when positive and negative terms occur with almost equal frequency in the sampling process. Before a sufficient amount of data has been collected, the error may thus substantially exceed the absolute value of the sampled integrals. The same applies to the error of the fitted spline. Fits where the noise substantially exceeds the signal are unsuitable for further data analysis. To identify this problem, the algorithm checks---at all bin levels---whether the sampled data are consistent with zero before the start of the fitting process.  The data is deemed \textit{certainly inconsistent} with zero only if at least one of the following conditions is met: (i) at any individual level $n$, the value of $\chi^2_n/\tilde{n}$ for the zero function exceeds 1 by at least 4 standard deviations ($4\sqrt{2/\tilde{n}}$), or (ii) at any two individual levels, $\chi^2_n/\tilde{n}$ exceeds 1 by at least $3\sqrt{2/\tilde{n}}$, respectively, or (iii) $\chi^2_n/\tilde{n}$ exceeds 1 by at least $3\sqrt{2/\tilde{n}}$ at one level, and by at least $2\sqrt{2/\tilde{n}}$ on two other levels, respectively, or (iv) at any four individual levels, $\chi^2_n/\tilde{n}$ exceeds 1 by at least $2\sqrt{2/\tilde{n}}$, respectively. The probability for this to occur as a statistical fluctuation is lower than $10^{-4}$. Unless one of these conditions is met, the algorithm can be set to not continue with the fitting procedure until more data has been collected. The conditions are chosen to be strict, to minimize the risk of continuing the analysis with a noisy fit.

An alternative to this procedure is to consider the evolution of the fitted spline with accumulated statistics,  accepting the spline only when it ceases to change substantially over Monte Carlo time,
\begin{equation}
\int \left|\tilde{f}_N(x) - \tilde{f}_{2N}(x)\right|dx < \alpha \int | \tilde{f}_{2N}(x) | \, dx ,
\end{equation}
where $\tilde{f}_N(x)$ is the BHM spline obtained from $N$ sampled data points, $\alpha$ is the acceptance threshold and the integration goes over the entire domain. Typical values for $\alpha$ lie approximately between $0.2$ and $0.5$.  

\begin{figure*}[htp]
\includegraphics[width=0.49\textwidth]{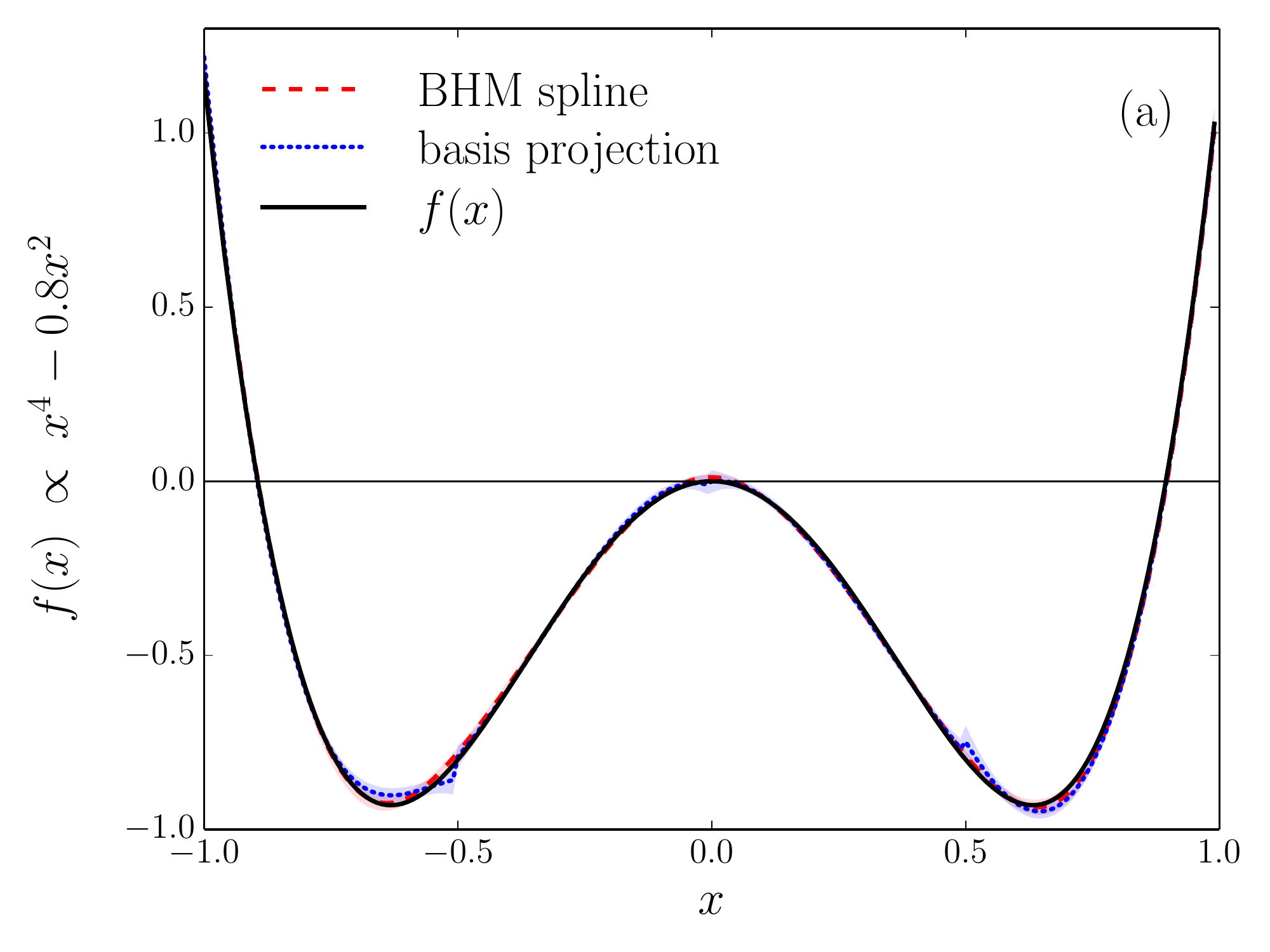}\hfill
\includegraphics[width=0.49\textwidth]{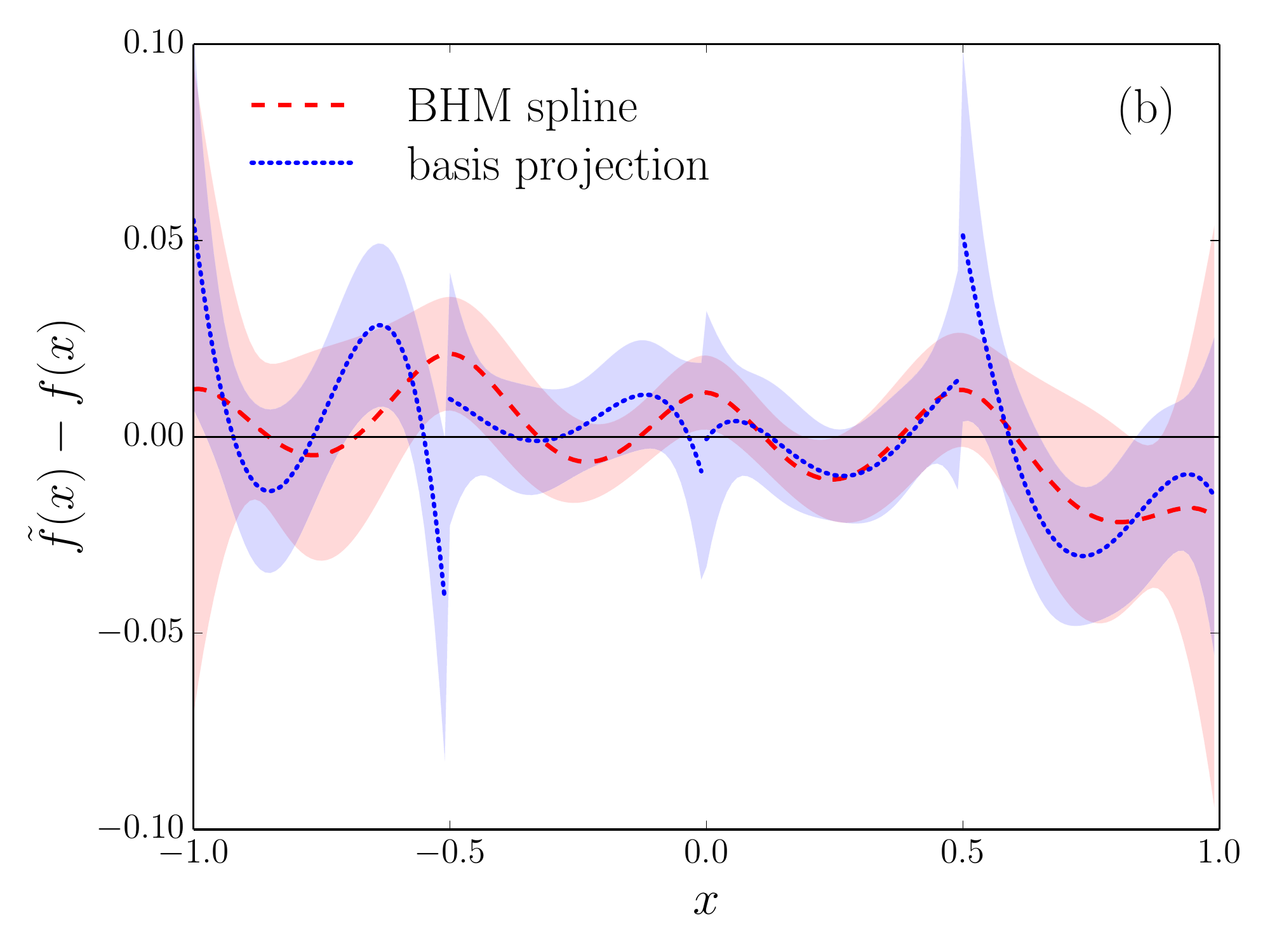}
\caption{\label{fig:hattest}Quartic polynomial test function. The BHM fit (red dashed line with error band) in comparison with the basis projection method (blue dotted line with error band). The left panel (a) shows the function and the fits, while the right panel (b) shows the difference between the fitted and the true function for both methods.}
\end{figure*}
\begin{figure*}[htp]
\includegraphics[width=0.49\textwidth]{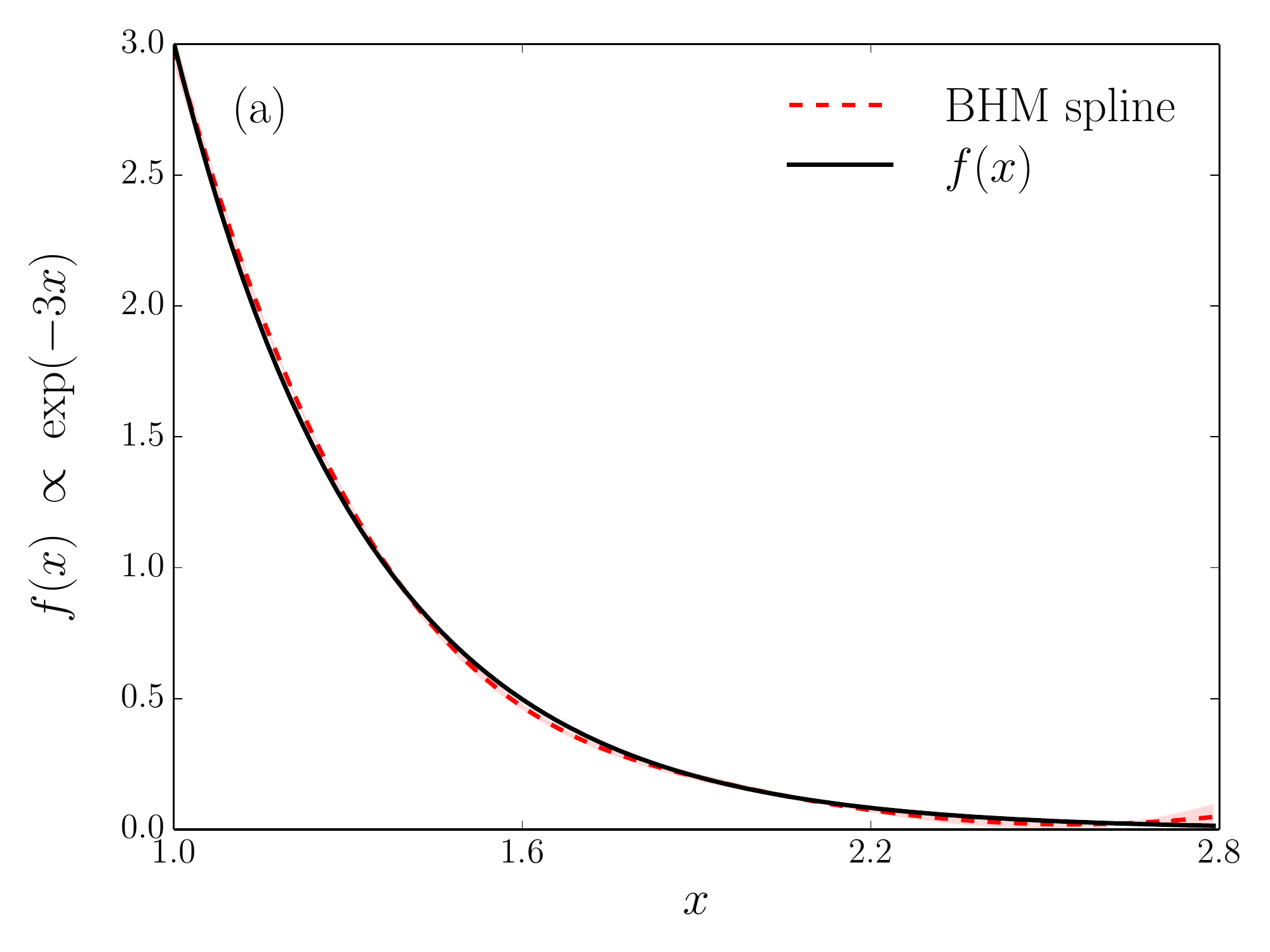}\hfill
\includegraphics[width=0.49\textwidth]{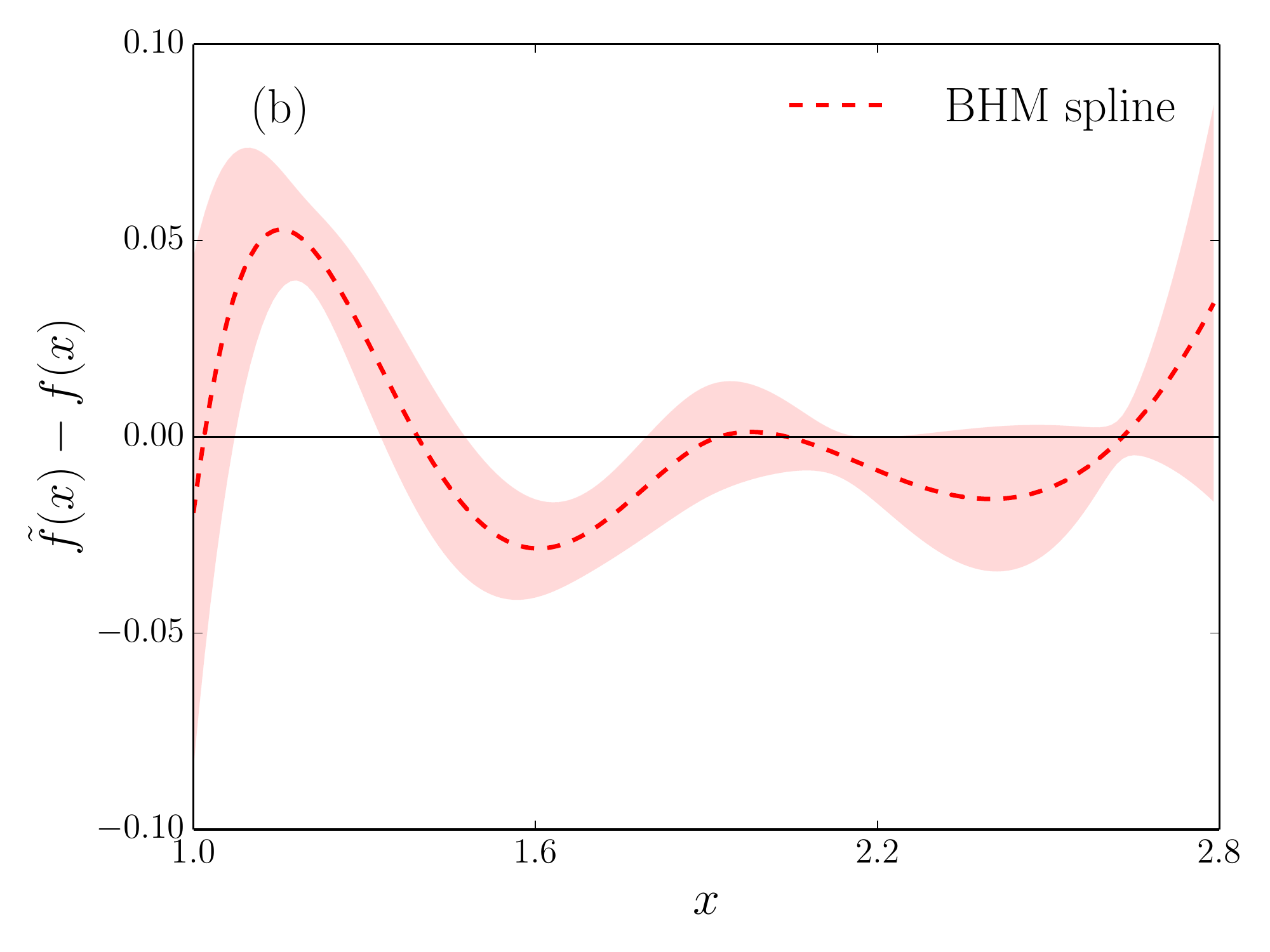}
\caption{\label{fig:exptest}BHM fit with error band of an exponential test function. The left panel (a) shows the function and the fit, while the right panel (b) shows the difference between the fitted and the true function.}
\end{figure*}
\begin{figure}[htp]
\includegraphics[width=\columnwidth]{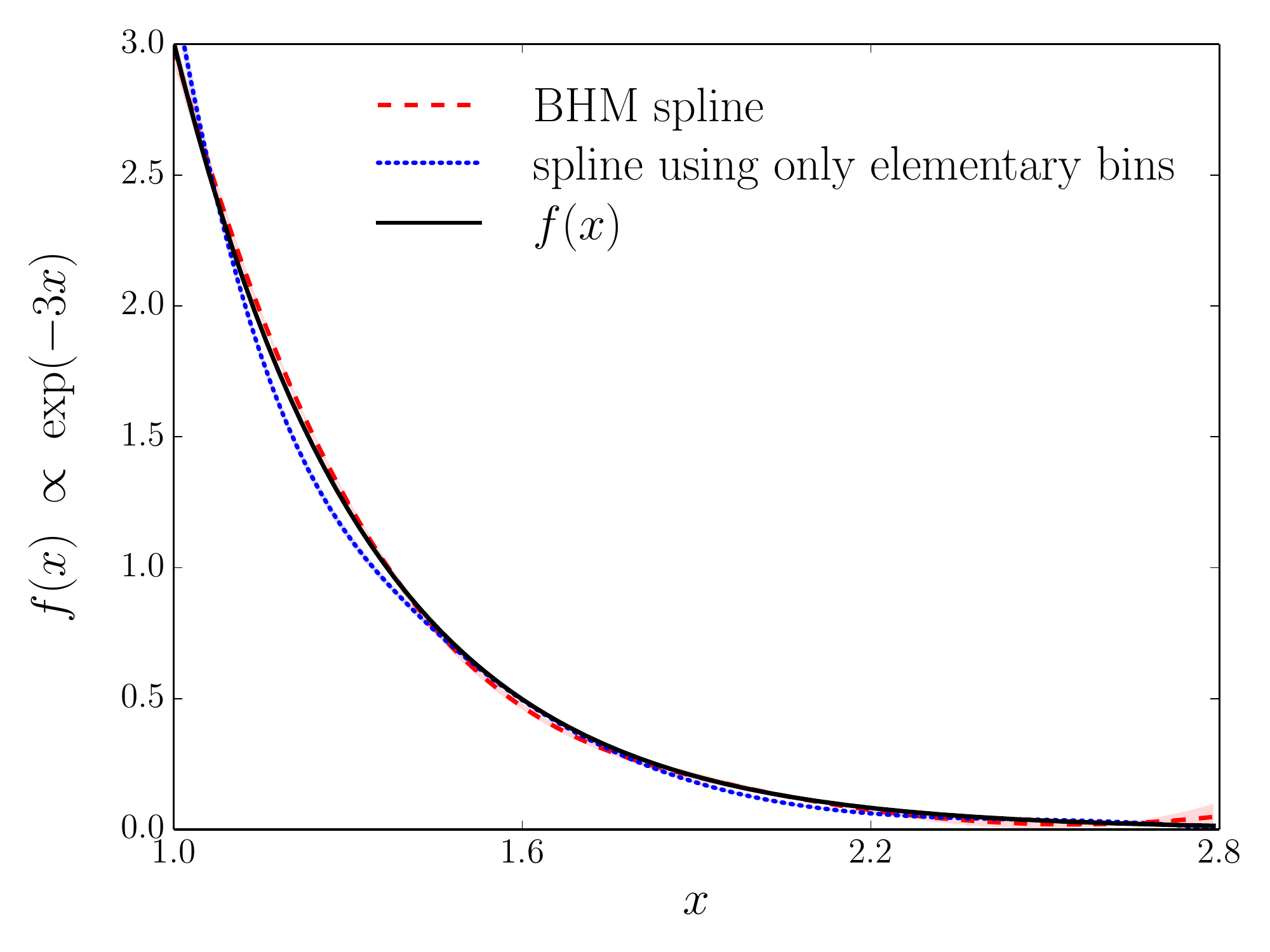}
\caption{\label{fig:exptest2}Exponential test function. The BHM fit (red dashed line with error band) in comparison with the fit using only elementary bins (blue dotted line with error band). The latter does not properly resolve all sampled integrals of the test function.}
\end{figure}
\begin{figure*}[htp]
\includegraphics[width=0.49\textwidth]{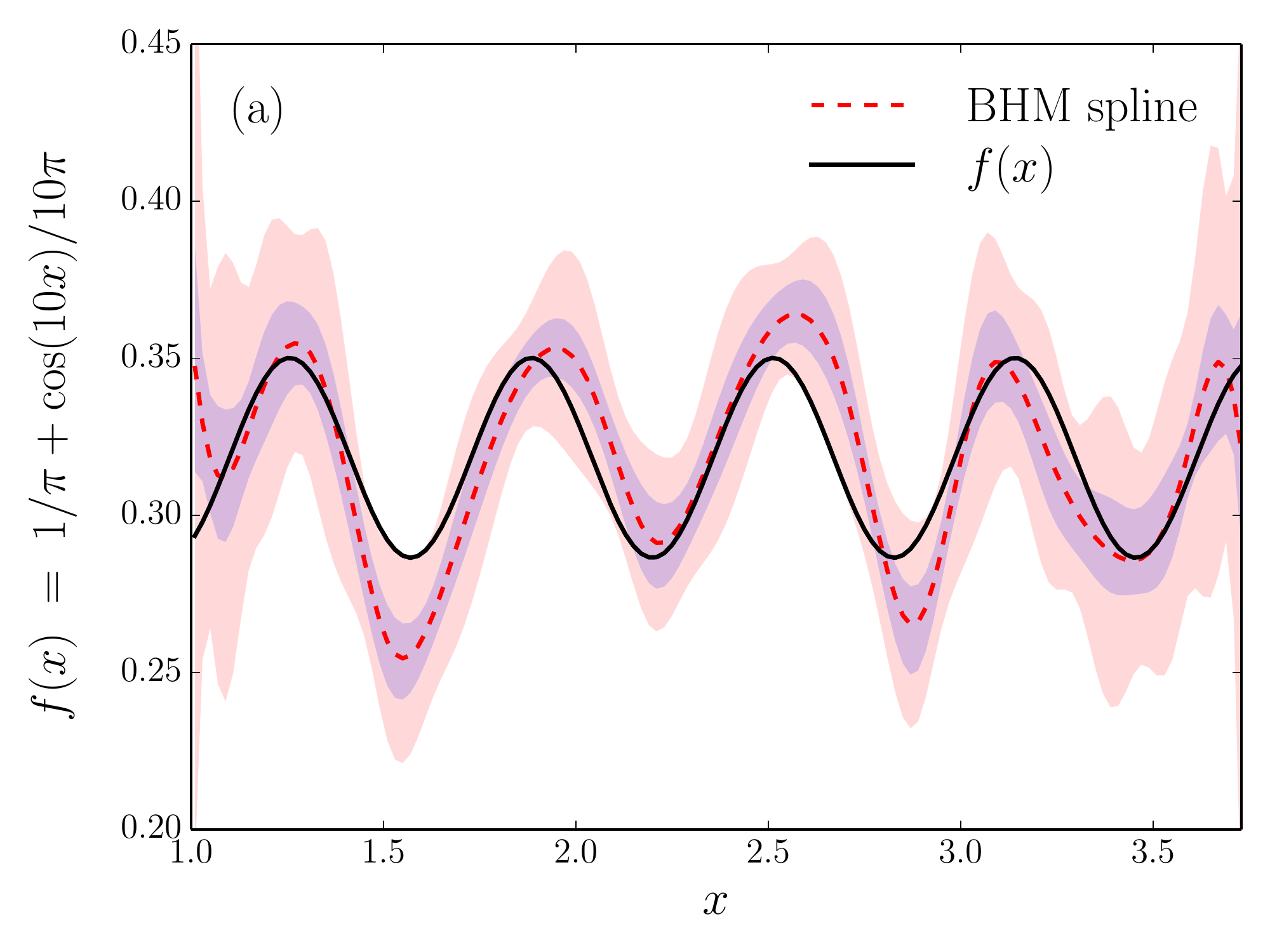}\hfill
\includegraphics[width=0.49\textwidth]{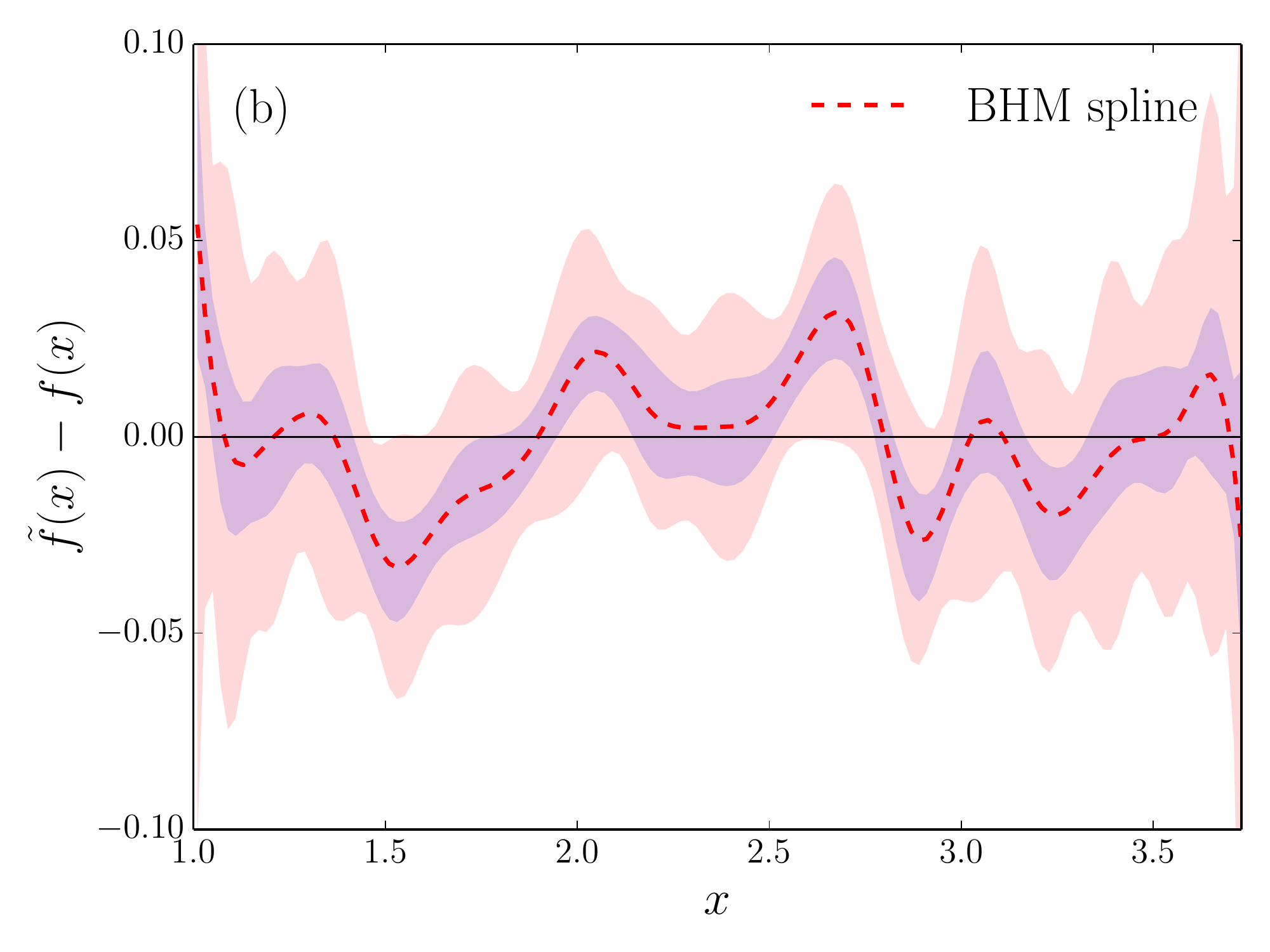}
\includegraphics[width=0.49\textwidth]{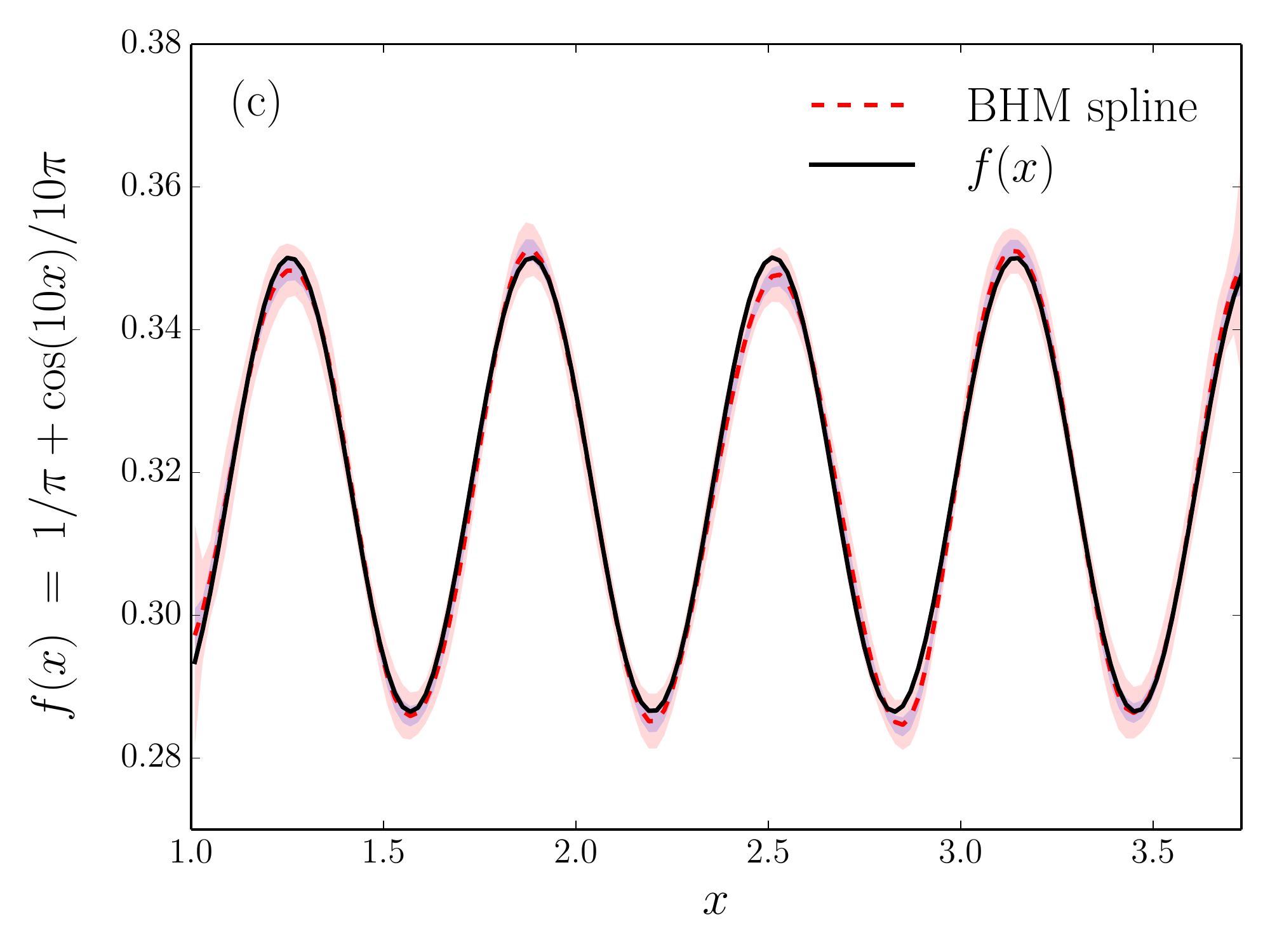}\hfill
\includegraphics[width=0.49\textwidth]{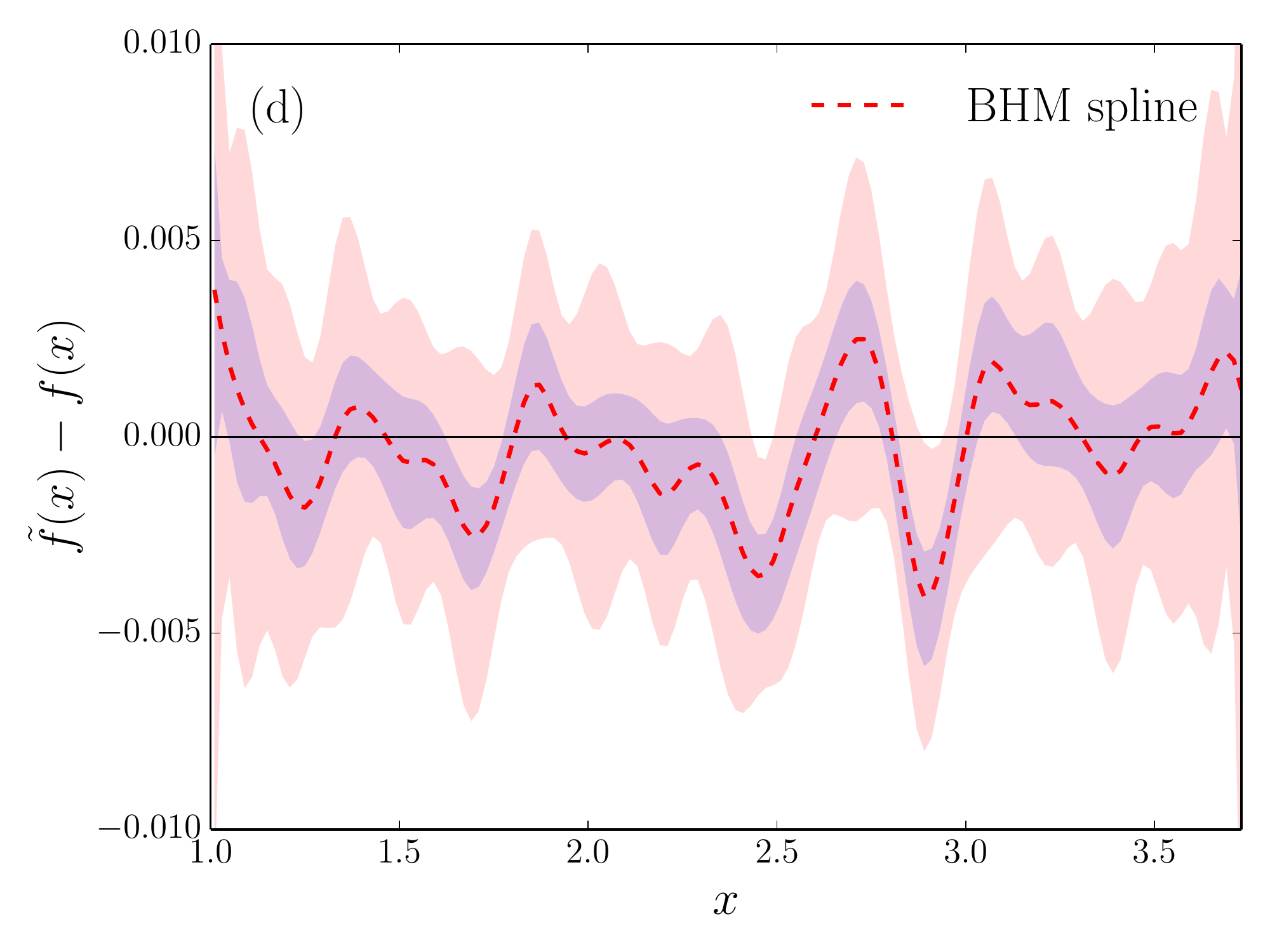}
\caption{\label{fig:costest}Oscillating test function using $10^4$ sampled points (top panels (a) and (b)) and $10^6$ sampled points (bottom panels (c) and (d)). The BHM fit (red dashed line) is shown with two error estimates on the spline coefficients: errors calculated using Eq.~\eqref{eq:coverror} (red band) and errors calculated using bootstrap (blue band). Bootstrap provides a tighter bound on the errors in this example. The left panels, (a) and (c), show the function and the fit, while the right panels, (b) and (d), show the difference between the fitted and the true function.}
\end{figure*}
\begin{figure}[htp]
\includegraphics[width=\columnwidth]{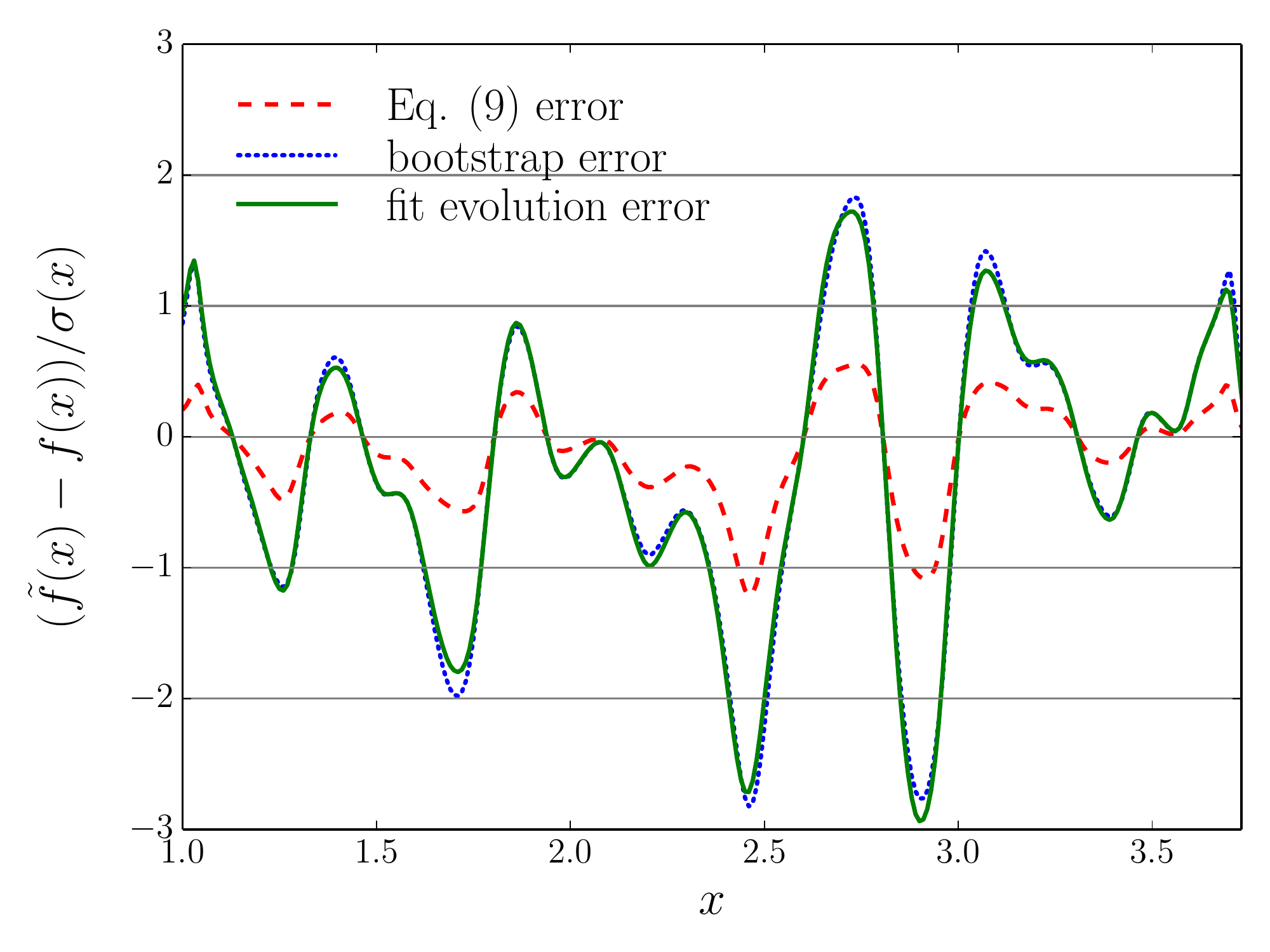}
\caption{\label{fig:coserr}Oscillating test function. Difference between the fitted and the true function normalized by one standard error obtained with different methods.}
\end{figure}
\begin{figure*}[htp]
\begin{tikzpicture}
    \node[anchor=south west,inner sep=0] at (0,0) {\includegraphics[width=\textwidth]{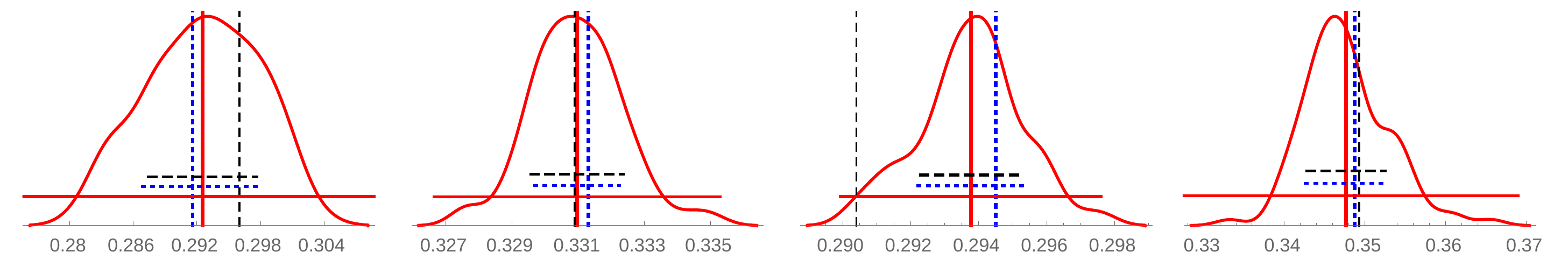}};
    \draw (0.5,2.8) node {(a)} (5,2.8) node {(b)} (9.3,2.8) node {(c)} (13.8,2.8) node {(d)};
\end{tikzpicture}
\caption{\label{fig:cossyst}Smoothed histogram of $\tilde{f}(x)$ values for the oscillating test function at $x=1,\,2,\,2.9,\,3.74$ (from left to right). The vertical lines are: histogram mean (red solid line); true function value (blue dotted line); the particular $\tilde{f}(x)$ value from the simulation shown in the bottom panel of Fig.~\ref{fig:costest} (black dashed line). The latter is shown for reference. The horizontal lines are: the $\pm1\sigma$ interval around the histogram mean, where $\sigma$ is the median standard error from Eq.~\eqref{eq:coverror} (red solid line); the $\pm1\sigma$ interval around the histogram mean, where $\sigma$ is the robust standard error (blue dotted line); the $\pm1\sigma$ interval around the histogram mean, where $\sigma$ is the typical bootstrap error, taken from the example shown in the bottom panel of Fig.~\ref{fig:costest} (black dashed line).}
\end{figure*}
\begin{figure*}[htp]
\includegraphics[width=0.49\textwidth]{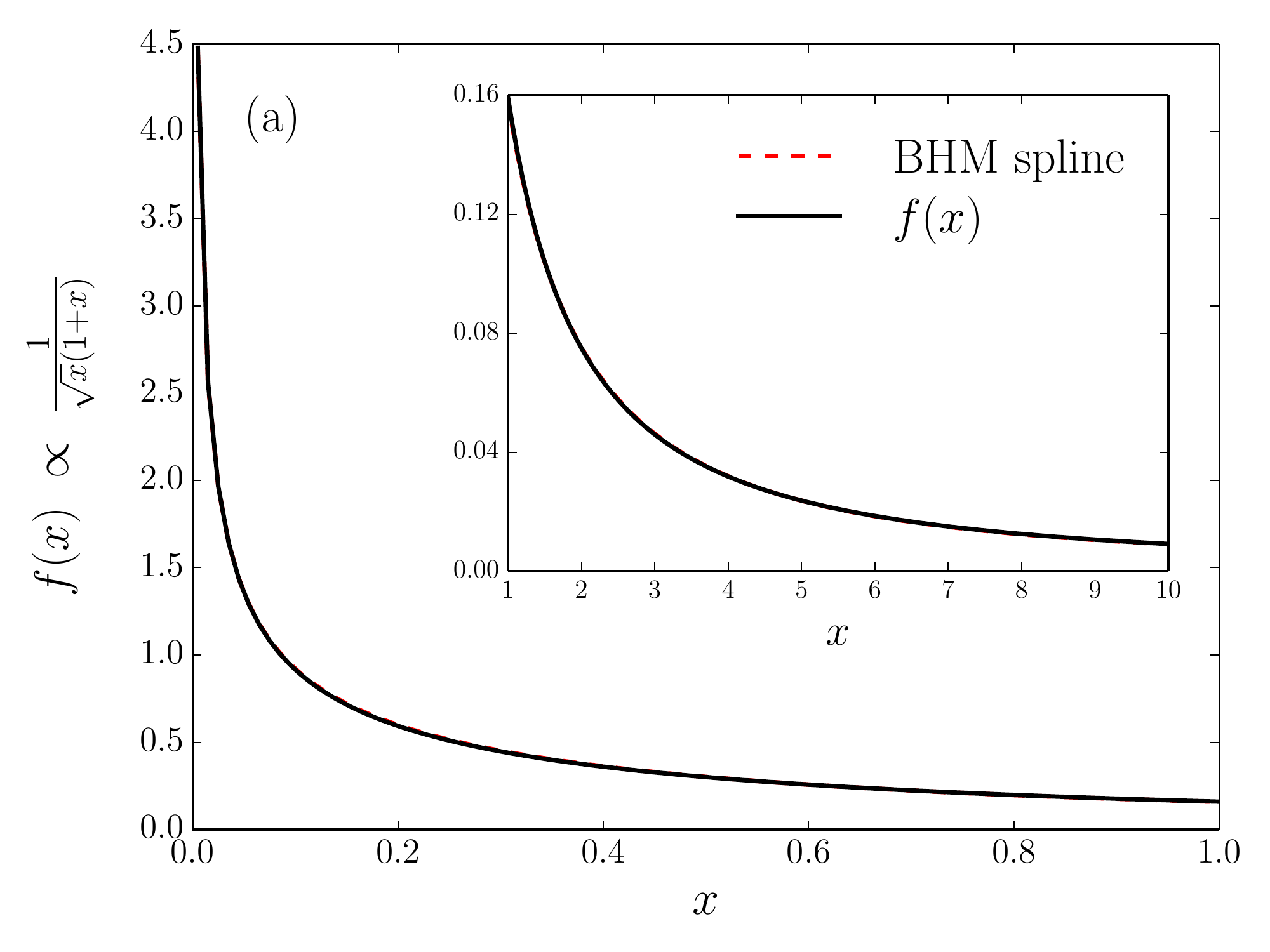}\hfill
\includegraphics[width=0.49\textwidth]{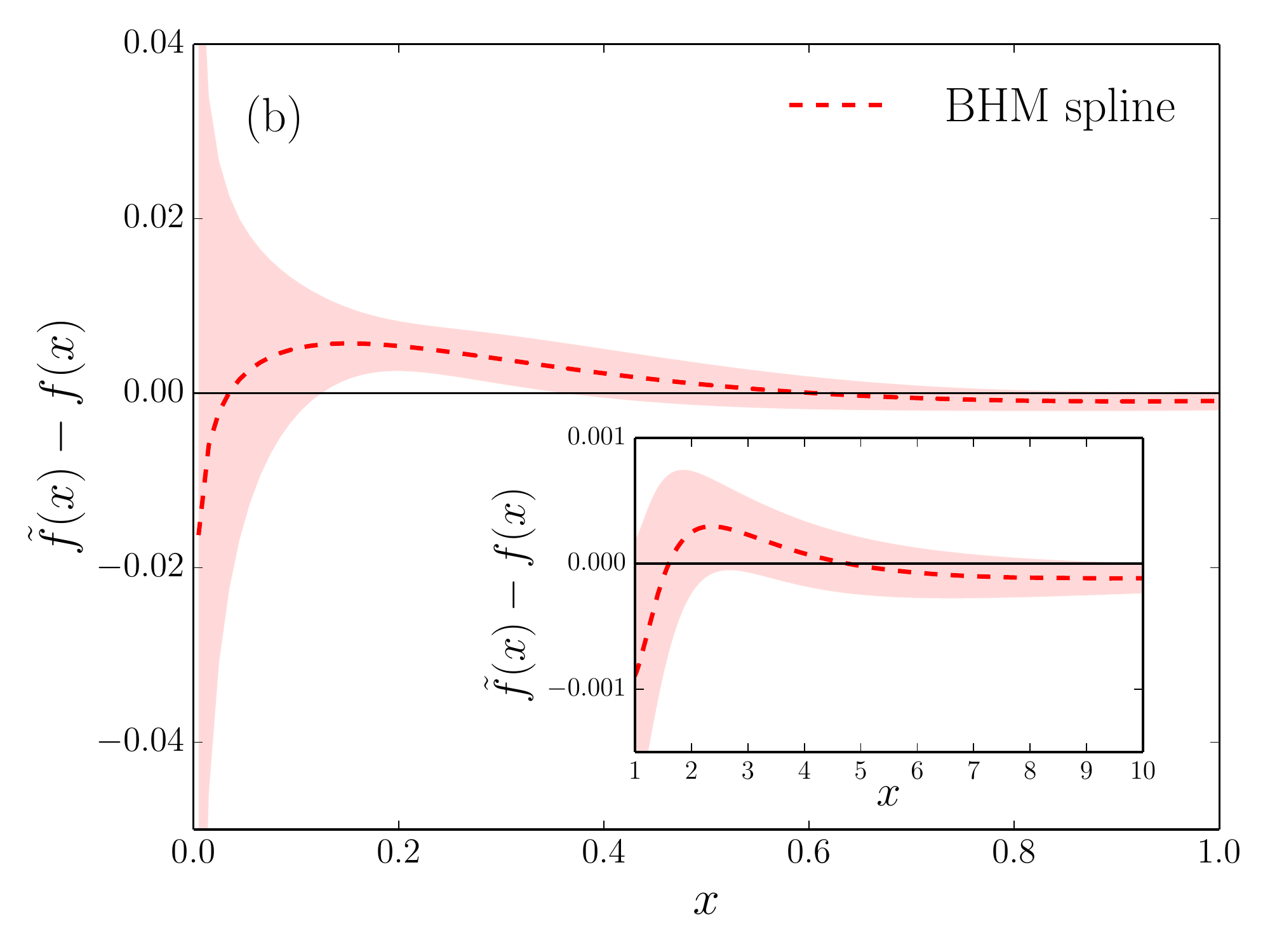}
\caption{\label{fig:divtest}Sampling a divergent function defined on a semi-infinite domain using the BHM with appropriate transforms. The left panel (a) shows the function and the fit, while the right panel (b) shows the difference between the fitted and the true function.}
\end{figure*}
\begin{figure*}[htp]
\includegraphics[width=0.49\textwidth]{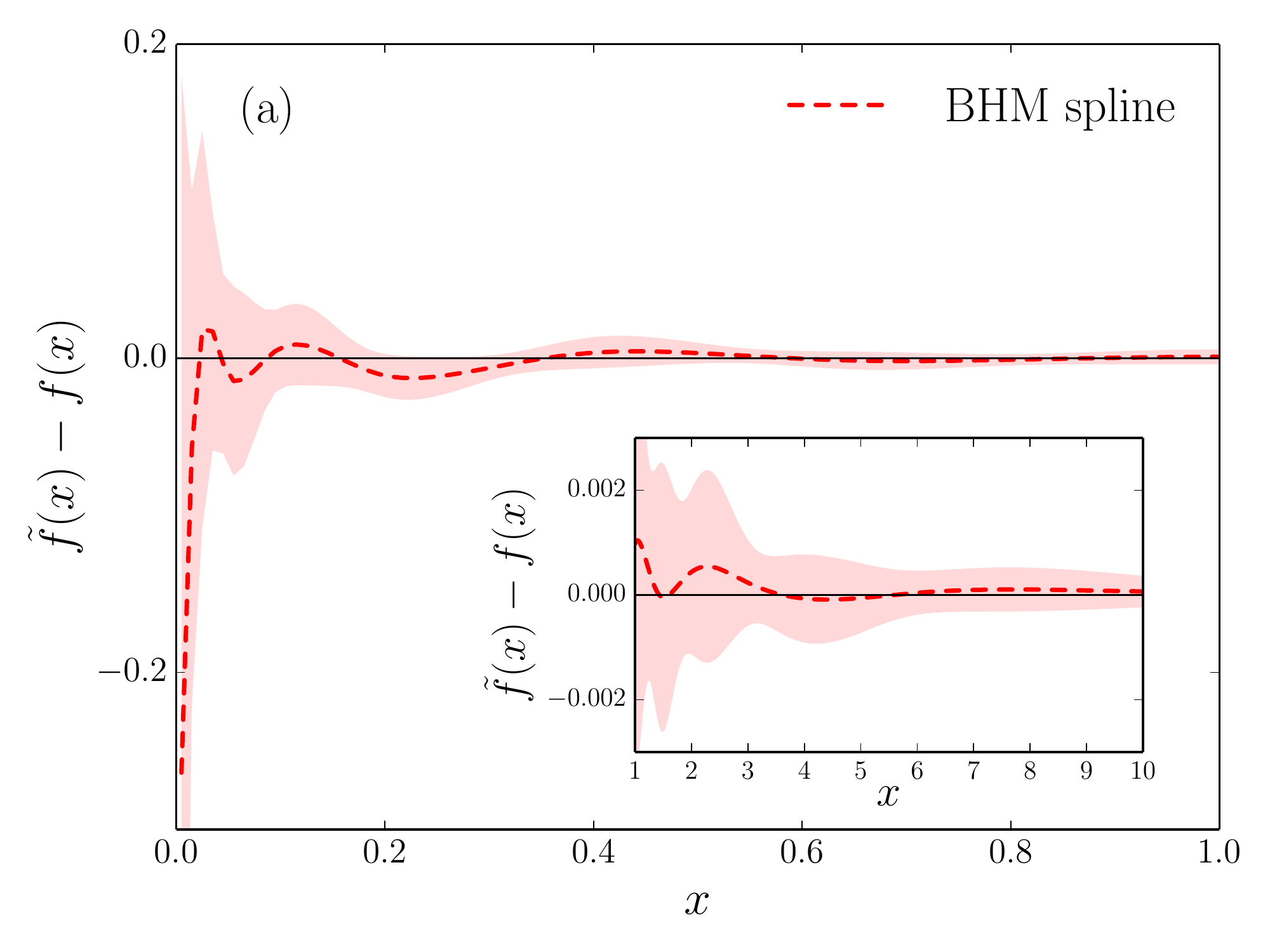}\hfill
\includegraphics[width=0.49\textwidth]{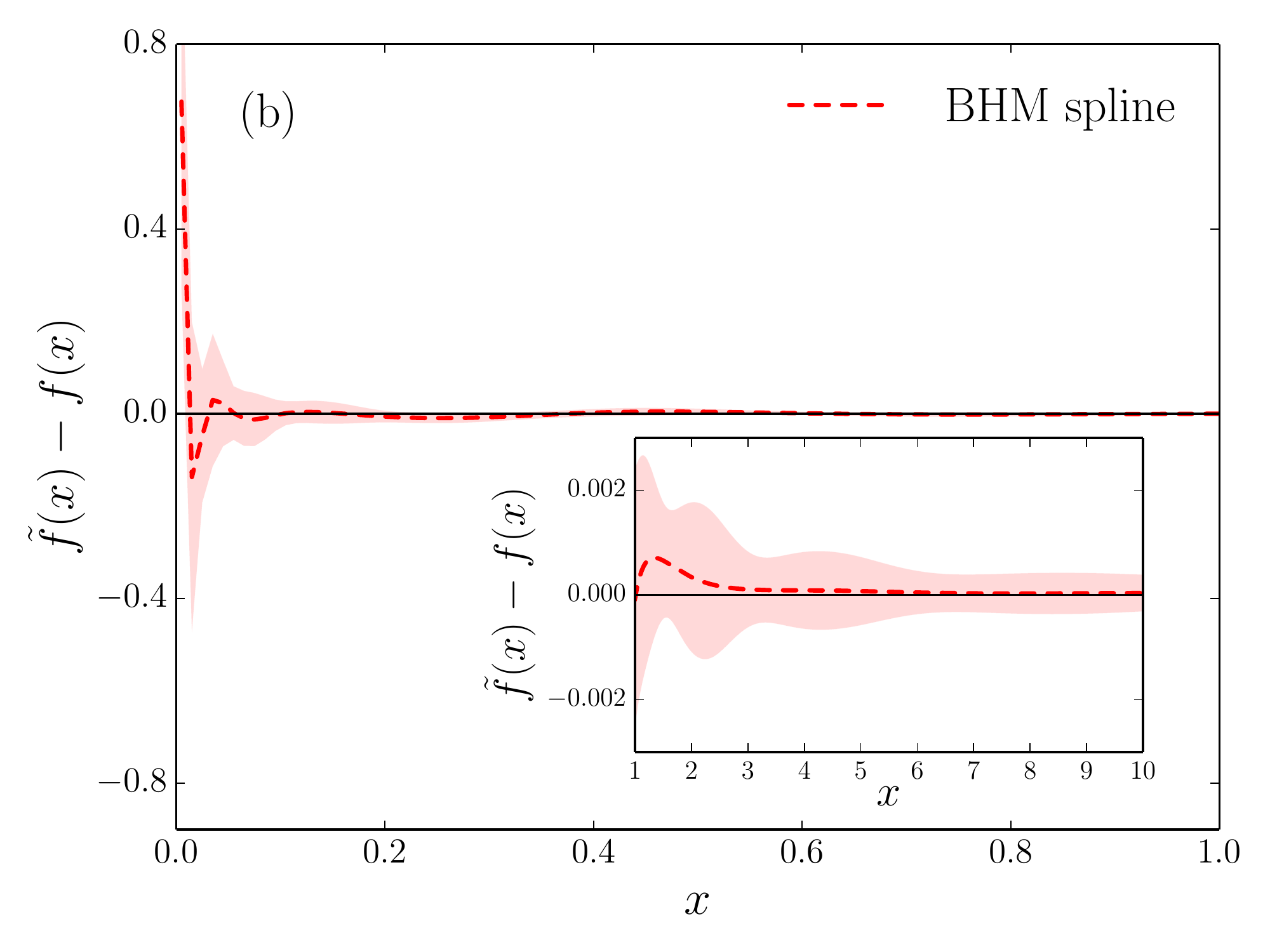}
\caption{\label{fig:divtestappr}Divergent test function $f(x)\propto x^{-0.5}(1+x)^{-1}$ with an $x^{-0.5}$ divergence at $x=0$. Data is sampled with a compensation for the divergence that overestimates (left panel (a)) and underestimates (right panel (b)) the power of the divergence by using weighting factors of $x^{0.8}$ and $x^{0.2}$, respectively. The plots show the difference between the BHM fit and the true function.}
\end{figure*}
\begin{figure*}[htp]
\includegraphics[width=0.49\textwidth]{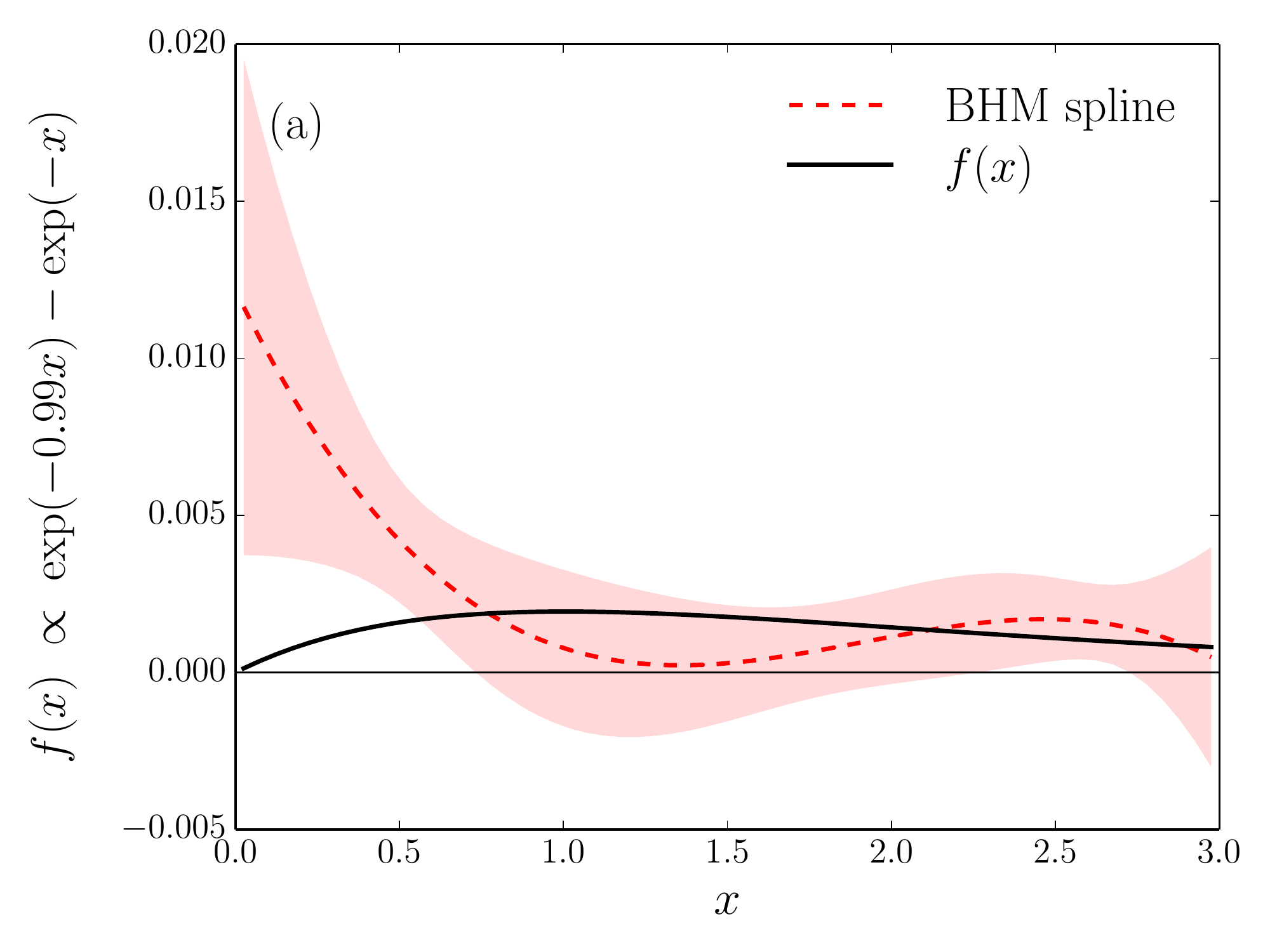}\hfill
\includegraphics[width=0.49\textwidth]{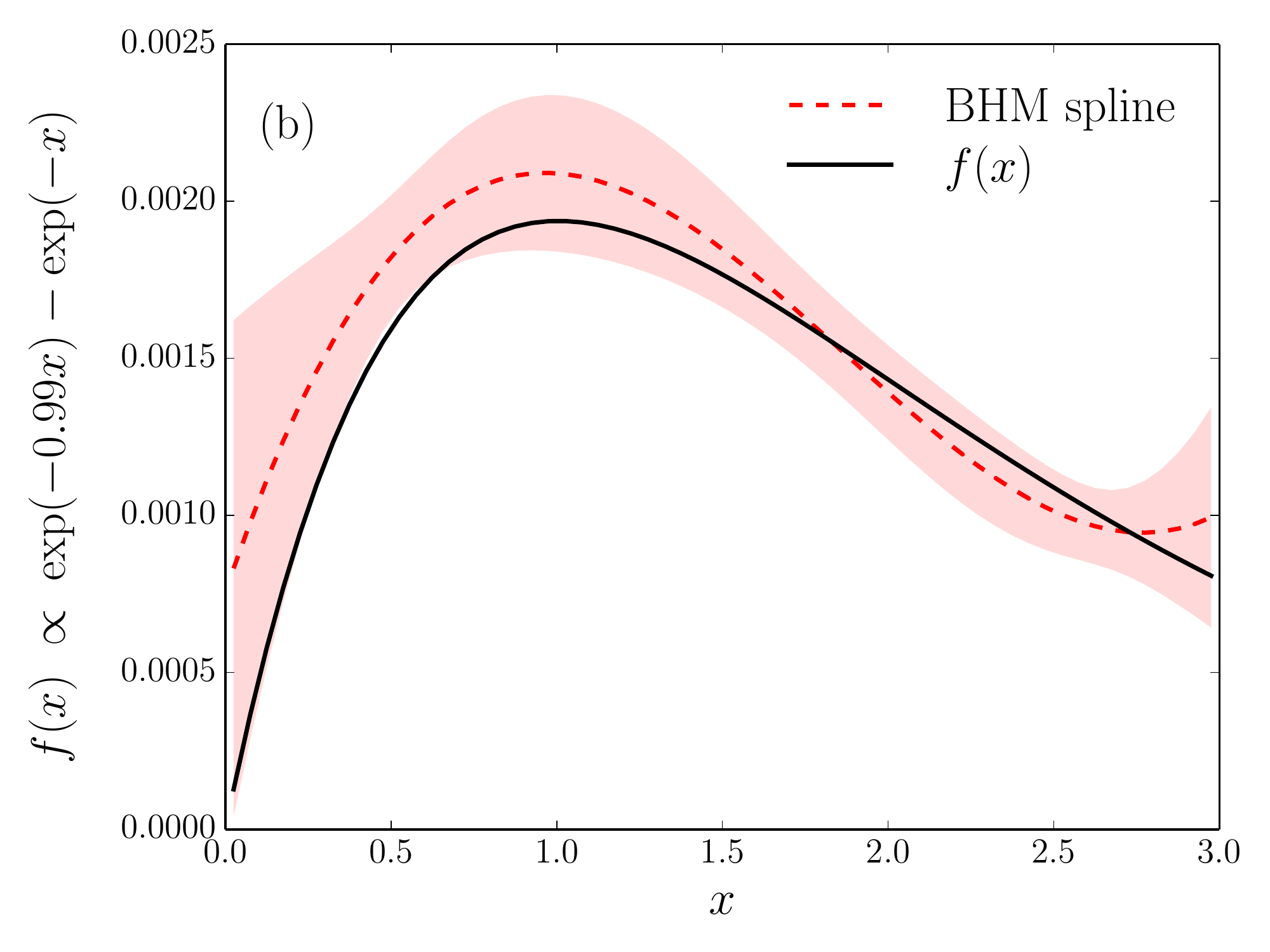}
\caption{\label{fig:signtest}Sampling with a severe sign problem using $10^5$ sampled points (left panel (a)) and $10^7$ sampled points (right panel (b)). The sampled integrals used for the fit in the left panel are consistent with zero, which is correctly captured by the check presented in Sec.~\ref{sec:signproblem}. In this case, the displayed BHM spline would not be used for further data analysis.}
\end{figure*}
\begin{figure*}[htp]
\includegraphics[width=0.49\textwidth]{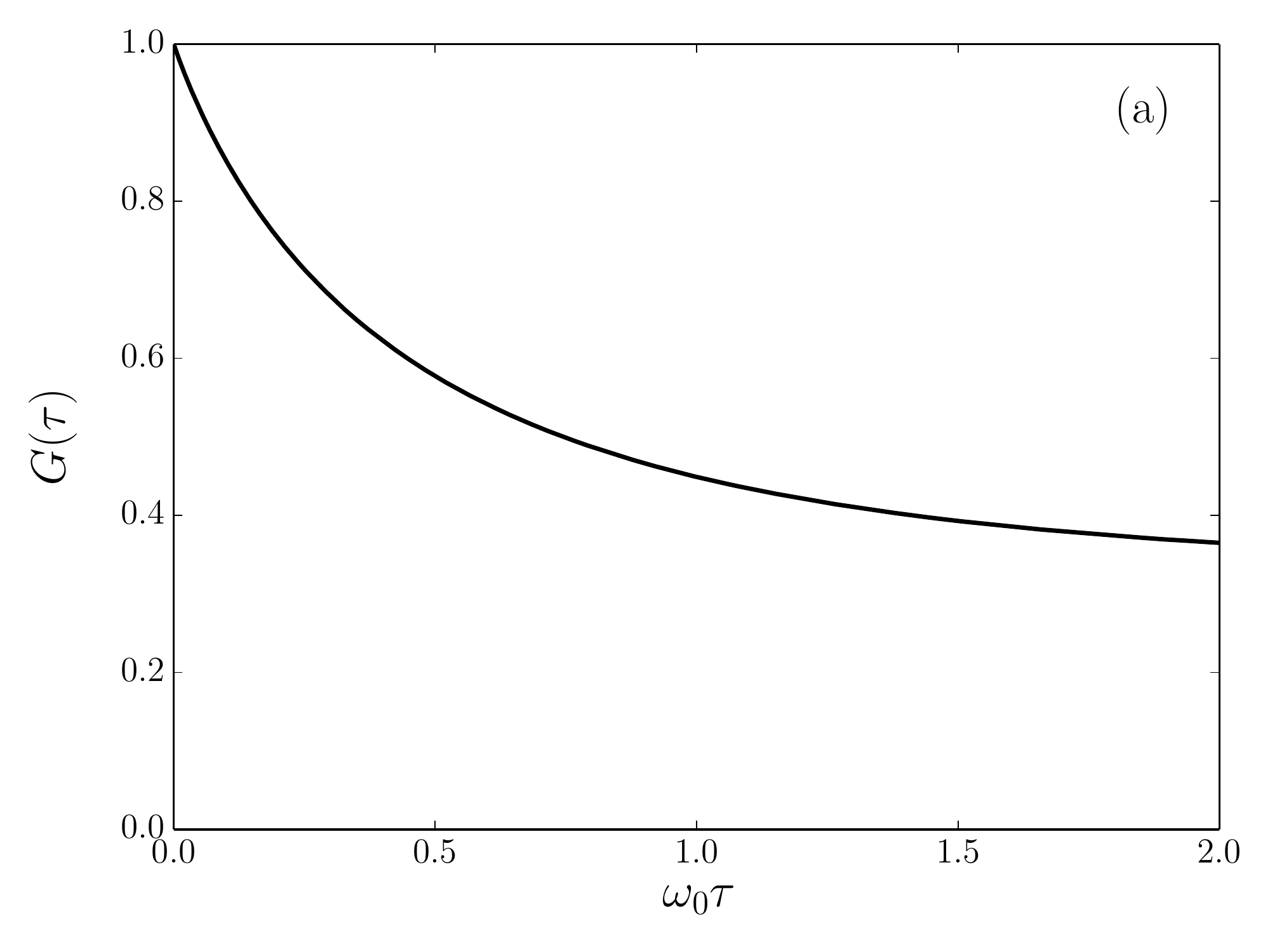}\hfill
\includegraphics[width=0.49\textwidth]{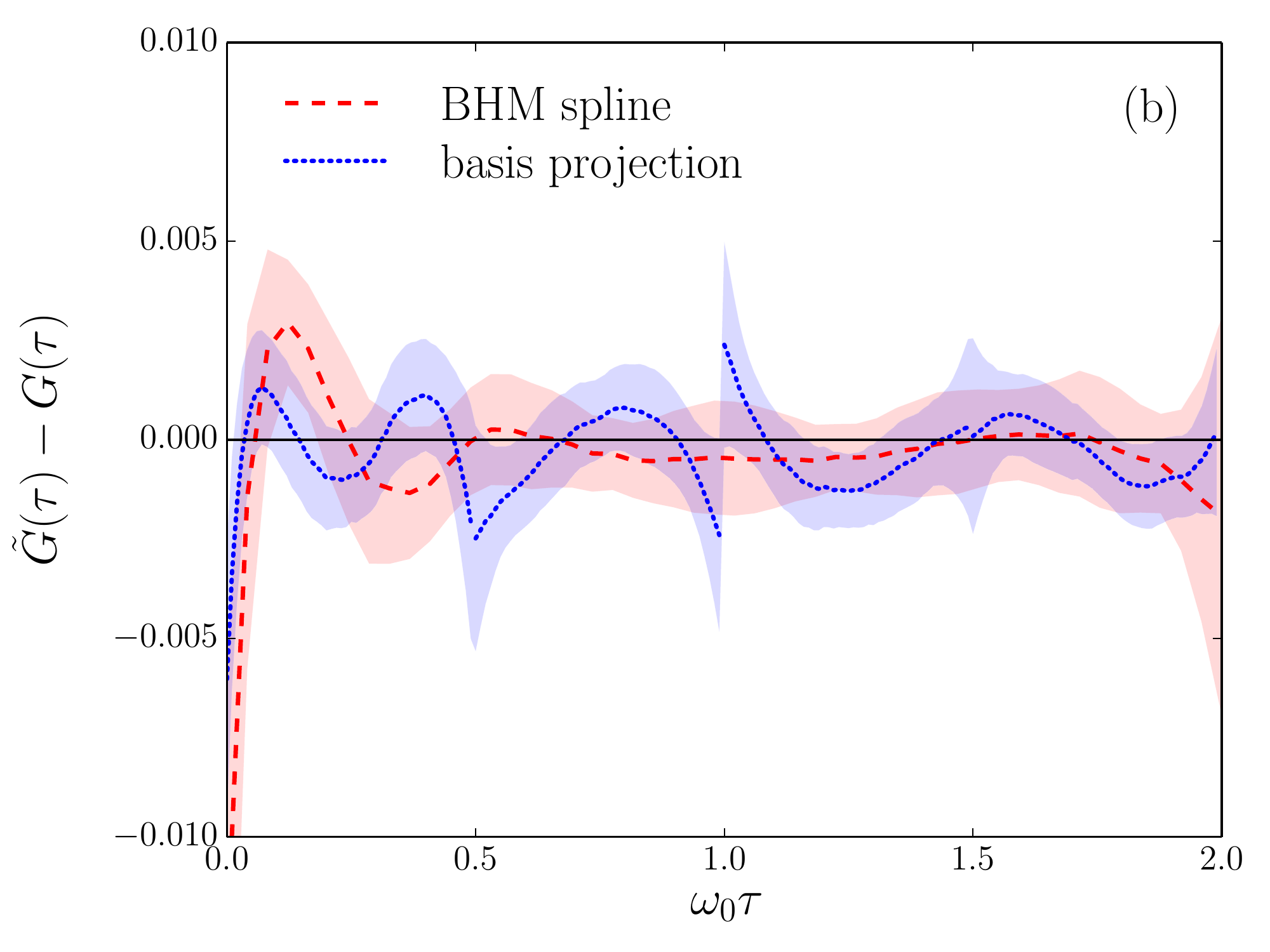}
\caption{\label{fig:green}Diagrammatic Monte Carlo sampling of the Fr\"ohlich polaron Green's function. The left panel (a) shows a very precise numerical calculation of the function, while the right panel (b) shows the difference between the fitted and the reference function for the BHM and the basis projection method.}
\end{figure*}

\section{Numerical tests}
\label{sec:tests}
We present tests of the BHM for several types of distribution. Unless otherwise stated, for each test $10^4$ random samples were taken, to demonstrate the ability to restore the smooth distribution from a moderate sample. We always fit cubic splines. The displayed error bands on the splines were obtained with Eq.~\eqref{eq:coverror}.

For comparison, we sampled the same data points also with the basis projection method using a polynomial basis up to cubic order. The binning for basis projection sampling was retrospectively chosen to be the same as the interval division for the spline that was determined by the bin hierarchy algorithm. This is a strict test, since in practice the optimal binning for the projection method is not known in advance. For each of the tests we observed that the accuracy of the bin hierarchy fit was at least as good as with the basis projection method, proving the superiority of BHM due to {\it at least} its efficiency and the smoothness of $\tilde{f}(x)$. We display the basis projection results for the first example of a polynomial distribution and omit them in the following examples to avoid overcrowding the plots.

\subsection{Polynomial}
The distribution sampled on the interval $[1,2.8]$ is 
\begin{equation}
f(x)\, \propto\,  1 - 3x/2 + 2x^2 - x^3/2.
\end{equation}
Since the distribution is a cubic polynomial, the algorithm should stop after fitting one polynomial on the entire domain. This is indeed the case. Figure~\ref{fig:poltest} shows the BHM fit in comparison with the result obtained using the basis projection method.  In this particular case, the basis projection method provides the most accurate estimate on the function by definition, since we are projecting on a function that has the same form as $f(x)$ and thus cover the whole interval $[1,2.8]$ without introducing any systematic error. Nevertheless, BHM produces a result comparable in accuracy. This example demonstrates that the BHM fit competes with the basis projection method even when the latter is known to be optimal.

We also perform tests with a higher order polynomial. The function sampled on the interval $[-1,1]$ is 
\begin{equation}
f(x)\, \propto \, x^4-0.8x^2.
\end{equation}
Since $f(x)$ changes sign, we use $|f(x)|$ as the probability to generate $x$ and then sample ${\rm sign}[f(x)]$. The BHM provides an accurate smooth fit of this function. The result in comparison with the basis projection method is shown in Fig.~\ref{fig:hattest}. Four spline pieces were needed to resolve the function in this example. As in the previous example, the errors on the BHM spline and the basis projection are comparable in size.

\subsection{Decaying exponential}
The distribution sampled on the interval $[1,2.8]$ is 
\begin{equation}
f(x)\, \propto \, \exp(-3x).
\end{equation}
An exponentially decaying distribution is typical for many physical processes. In this example, the BHM produced an acceptable fit with two spline pieces on the intervals $[1,1.9]$ and $[1.9,2.8]$. Figure~\ref{fig:exptest} shows the best fit, which agrees well with the original function.

To demonstrate the importance of including combinations of bins into the fit, we also compare our result with the best spline obtained by fitting elementary bins only (those elementary bins that did not contain enough data points for sensible statistics were coarse grained until the minimal size that can be used for statistical analysis was reached). Figure~\ref{fig:exptest2} compares the two fits. The fit using only elementary bins deviates strongly from $f(x)$ on the first interval $[1,1.9]$. In particular, it is apparent that the integrals over this interval and over the entire domain were not properly captured.

\subsection{Oscillating function}
\label{sec:osc}
The distribution sampled on the interval $[1,\pi+0.6]$ is 
\begin{equation}
f(x)\, \propto \, 10+\cos(10x).
\end{equation}
The challenge is to resolve the periodic oscillations within the accuracy of the sampled data. For a small sample, we expect a flat fit reproducing the average of the function. As we collect more data the oscillations should be resolved. We show the bin hierarchy fit for a small sample of $10^4$ points and for a larger sample of $10^6$ points in Fig.~\ref{fig:costest}. The fit reproduces the original distribution well within error bars. Many spline pieces are needed to resolve the structure (in the examples shown: 15--16 intervals) and Eq.~\eqref{eq:coverror} overestimates the error on the spline coefficients. In Fig.~\ref{fig:costest} we show the error band obtained from Eq.~\eqref{eq:coverror} as well as the one obtained using bootstrap.

In Fig.~\ref{fig:coserr} we explicitly compare the error estimates obtained with Eq.~\eqref{eq:coverror}, bootstrap, and fit evolution. The bootstrap and fit evolution errors are nearly the same for all $x$. In this example, for bootstrap and fit evolution errors, the deviation $\tilde{f}(x)-f(x)$ exceeds $1\sigma$ on roughly $30\%$ of the domain and $2\sigma$ on roughly $7\%$ of the domain, which is comparable to the Gaussian case. For the Eq.~\eqref{eq:coverror} error, the deviation almost never exceeds $1\sigma$. We also present a robust error analysis based on $100$ independent samples with $10^6$ points each. The histogram of $\tilde{f}(x)$ values is shown in Fig.~\ref{fig:cossyst} for several values of $x$. For reference we indicate the specific value of $\tilde{f}(x)$ from the example shown in Fig.~\ref{fig:costest}, to illustrate for which $x$ the fit values in this example are statistical outliers, and for which they are typical. We also show the median of the Eq.~\eqref{eq:coverror} errors and a representative bootstrap error (from the example shown in Fig.~\ref{fig:costest}). Indeed, the size of the bootstrap and fit evolution errors is very close to the robust estimate, while the error from Eq.~\eqref{eq:coverror} is too large.

\subsection{Divergent function on a semi-infinite domain}
The distribution sampled on the interval $[0,\infty)$ is 
\begin{equation}
f(x)\, \propto \, \frac{1}{\sqrt{x}(1+x)}.
\end{equation}
The basis projection method has the advantage that the $1/\sqrt{x}$ divergence at $x=0$ can be incorporated directly into the basis, and that the asymptotic behavior at large $x$ can be resolved using a semi-infinite bin. We present a setup that achieves the same in the BHM framework. The divergence can be avoided by weighting all sampled values with $\sqrt{x}$ and recovering the original distribution at the end by dividing the BHM spline by $\sqrt{x}$. The semi-infinite domain can be mapped onto a finite interval for instance through a transform such as $y(x)=2\arctan(x)/\pi$ (used in this example) or $y(x)=1-\exp(-x)$. For each $x$ generated according to the distribution $f(x)$, the value $y(x)\in[0,1]$ is calculated and sampling is then performed into the corresponding $y$-histogram. To recover the correct function, the sampled values need to be scaled by $1/x'(y)$.

Figure~\ref{fig:divtest} shows the BHM fit transformed back into the original domain. Due to the large domain, $10^5$ sampling points were used for this example. The fit agrees well with the original distribution within error bars. In this example, the BHM produced an acceptable fit on one spline interval.

In order to compensate for the divergence, it is assumed that its location and type are known in advance. Often it is sufficient to know the approximate properties of the divergence. To demonstrate this, Fig.~\ref{fig:divtestappr} shows BHM fits of data sampled with a weighting factor that slightly overestimates or underestimates the power of the divergence (the mapping of the semi-infinite domain onto a finite interval is the same as before). Despite the wrong scaling, the fits still agree with the original distribution, but the errors on the fits are larger. It is better to overestimate the power, since in this case the divergence is still fully compensated.

\subsection{Sampling with a severe sign problem}
We demonstrate the performance of the algorithm in the presence of a severe sign problem by generating data on the interval $[0,3]$ with the distribution given by
\begin{equation}
f(x)\propto \exp(-0.99x)-\exp(-x)\equiv f_1(x)-f_{-1}(x).
\end{equation}
The values of $x$ are generated via a Monte Carlo Markov chain process with two types of updates: switching between two ``sectors" corresponding to $f_1$ and $f_{-1}$, and varying the value of $x$ on the interval $[0,3]$ within the same sector. The update acceptance probabilities are given by the detailed balance equations,
\begin{eqnarray}
P_{f_{-\sigma}\rightarrow f_{\sigma}}&=&f_\sigma(x)/f_{-\sigma}(x)=\exp(\pm0.01x),\\
P_{x\rightarrow x'}&=&f_\sigma(x'-x).
\end{eqnarray}
The sign problem is due to the functions $f_1(x)$ and $f_{-1}(x)$ having almost the same magnitude but being sampled with opposite sign. This setup is a toy version of the typical situation arising in diagrammatic Monte Carlo for fermions.

Figure~\ref{fig:signtest} shows the BHM spline for $10^5$ and $10^7$ sampled points. In the former case, the data is compatible with zero, which is correctly captured by the algorithm presented in Sec.~\ref{sec:signproblem}. In this case, the fit should not be used, despite having acceptable $\chi^2$ (the fit is consistent with the true function within errors). As more data are gathered the algorithm begins to correctly reproduce the features of the function. In the examples shown, one spline piece was sufficient for an acceptable fit.

\subsection{Green's function of the Fr\"ohlich polaron}
We now apply the BHM to a physical problem---the diagrammatic Monte Carlo calculation of the zero momentum imaginary time Green's function,\cite{sompolaron} 
\begin{eqnarray}
G(\mathbf{k}=0,\tau)&=&\langle\textnormal{vac}|a_\mathbf{0}(\tau)a^\dagger_\mathbf{0}(0)|\textnormal{vac}\rangle,\\
a_\mathbf{k}(\tau)&=&e^{H\tau}a_\mathbf{k}e^{-H\tau}.
\end{eqnarray}
of the Fr\"ohlich polaron with the dimensionless coupling constant $\alpha=2$ and the chemical potential $\mu=-2.07\omega_0$, where $\omega_0$ is the (momentum-independent) phonon frequency. Here $|\textnormal{vac}\rangle$ is the vacuum state and $a_\mathbf{k}$ is the annihilation operator for an electron with momentum $\mathbf{k}$. The Fr\"ohlich Hamiltonian describes an electron coupled to a bath of phonons,\cite{froehlich}
\begin{eqnarray}
H&=&H_e+H_{ph}+H_{e-ph},\\
H_e&=&\sum_\mathbf{k}\frac{k^2}{2}a^\dagger_\mathbf{k}a_\mathbf{k},\ \ \ H_{ph}=\sum_\mathbf{q}\omega_0b^\dagger_\mathbf{q}b_\mathbf{q},\\
H_{e-ph}&=&\sum_{\mathbf{k},\mathbf{q}}\frac{i(2\sqrt{2}\alpha\pi)^{1/2}}{q}(b^\dagger_\mathbf{q}-b_{-\mathbf{q}})a^\dagger_\mathbf{k-q}a_\mathbf{k},
\end{eqnarray}
where $b_\mathbf{q}$ is the annihilation operator for a phonon with momentum $\mathbf{q}$.

The Green's function is a central quantity for the diagrammatic technique, from which other properties of the system can be obtained with appropriate analysis. As reference we use a very long Monte Carlo run with reweighting. This provides a reliable estimate of the Green's function with negligible errors (several orders of magnitude smaller than the errors of the sample used for the BHM and for basis projection sampling).

Figure~\ref{fig:green} shows the BHM fit and the result obtained using the basis projection method for approximately $2\cdot10^6$ sampled points. The BHM fit produced four spline pieces in this example and, as in the previous examples, the basis was retrospectively chosen to have the same interval division. It can be clearly seen that the BHM provides an accurate smooth fit of the Fr\"ohlich polaron Green's function, which agrees well with the reference. While the error bars on the BHM fit and the basis projection are comparable, the basis projection sampling is less efficient, since it requires $\mathcal{O}(m^2)$ operations during the sampling stage, where $m$ is the number of basis functions used. In this particular example with a polynomial basis up to cubic order, this corresponds to at least 16 times as many operations as needed during the sampling stage for BHM.

\section{Conclusions}
\label{sec:conclusions}
We have argued and demonstrated that the BHM yields an efficient, flexible, and fully automatized algorithm to restore smooth functions from their noisy integrals using all available information. The resulting fits are at least as accurate as the ones obtained using basis projection sampling, but with guaranteed smoothness at the knots. Sampling with the BHM is also computationally less expensive than using basis projections, which require many operations in each sampling step. Similar to the projection method onto a polynomial basis, the BHM spline is a piecewise polynomial function on several large intervals. The crucial technical advantage is that these intervals do not need to be fixed beforehand. A suitable division into intervals is found automatically by the algorithm and is adjusted over time, as more data points are collected. 

In the future we plan to extend the bin hierarchy algorithm to multivariate functions, since many relevant physical observables depend on several variables, such as time and momentum. Two, three and four dimensions are most relevant for physical applications.

\begin{acknowledgments}
We thank Chris Amey for helpful discussions. This work was supported by the Simons Collaboration on the Many Electron Problem and the National Science Foundation under the grant DMR-1720465. O.G.\ also acknowledges support by the US-Israel Binational Science Foundation (Grants 2014262 and 2016087).
\end{acknowledgments}

\bibliography{allbib}

\end{document}